\documentclass[11pt,prl,twocolumn,english,aps,reprint,superscriptaddress,notitlepage,nobibnote,preprintnumbers,showpacs]{revtex4-2}

\usepackage{graphicx}
\usepackage{dcolumn}
\usepackage{bm}
\usepackage{amsmath}
\usepackage{amssymb}
\usepackage{mathrsfs}
\usepackage[export]{adjustbox}
\usepackage{color}
\usepackage{xcolor}
\usepackage{autobreak}
\usepackage{multirow}
\usepackage{subfig}
\usepackage[vcentermath]{youngtab}
\usepackage{pifont}
\usepackage{extarrows}
\usepackage{makecell}
\usepackage{youngtab}
\usepackage{ytableau}
\usepackage{ulem} 
\usepackage{bbm}
\usepackage{diagbox}
\usepackage{slashed}
\usepackage{tikz}
\usepackage[compat=1.1.0]{tikz-feynhand}
\usetikzlibrary{snakes}
\usetikzlibrary{arrows}

\usepackage[toc]{appendix}
\usepackage[colorlinks=true,
linkcolor=blue,
urlcolor=red,
citecolor=red]{hyperref}

\usepackage[utf8]{inputenc}
\usepackage{times} 
\usepackage{mathptmx} 

\makeatletter
\newcommand{\xRightarrow}[2][]{\ext@arrow 0359\Rightarrowfill@{#1}{#2}}
\makeatother

\newcommand{\beq}{\begin {equation}}  
\newcommand{\eeq}{\end   {equation}} 
\newcommand{\bea}{\begin {eqnarray}} 
\newcommand{\eea}{\end   {eqnarray}}  
\newcommand{\baa}{\begin {array}   } 
\newcommand{\eaa}{\end   {array}   }     
\newcommand{\bit}{\begin {itemize} }
\newcommand{\eit}{\end   {itemize} }
\newcommand{\be }{\begin {equation}} 
\newcommand{\ee }{\end   {equation}}

\newcommand{\eq}[1]{\begin{equation}\begin{split} #1 \end{split}\end{equation}}
\newcommand\undermat[2]{
	\makebox[0.5pt][l]{$\smash{\underbrace{\phantom{%
					\begin{matrix}#2\end{matrix}}}_{ \let\scriptstyle\textstyle\text{\normalsize $#1$}}}$}#2}
\newcommand\overmat[2]{
	\makebox[-1pt][l]{$\smash{\overbrace{\phantom{%
					\begin{matrix}#2\end{matrix}}}^{ \let\scriptstyle\textstyle\text{\normalsize $#1$}}}$}#2}

\begin{document}

\title{Extended Poincar\'e Symmetry Dictates Massive Scattering Amplitudes}

\author{Yu-Han Ni}
\email{niyuhan@itp.ac.cn}
\affiliation{
CAS Key Laboratory of Theoretical Physics, Institute of Theoretical Physics, \\ Chinese Academy of Sciences, Beijing 100190, P.\ R.\ China}
\affiliation{
School of Physical Sciences, University of Chinese Academy of Sciences, Beijing 100049, P.\ R.\ China}

\author{Yi-Ning Wang}
\email{wangyining@itp.ac.cn}
\affiliation{
CAS Key Laboratory of Theoretical Physics, Institute of Theoretical Physics, \\ Chinese Academy of Sciences, Beijing 100190, P.\ R.\ China}
\affiliation{
School of Physical Sciences, University of Chinese Academy of Sciences, Beijing 100049, P.\ R.\ China}

\author{Chao Wu}
\email{wuch7@itp.ac.cn}
\affiliation{
CAS Key Laboratory of Theoretical Physics, Institute of Theoretical Physics, \\ Chinese Academy of Sciences, Beijing 100190, P.\ R.\ China}

\author{Jiang-Hao Yu}
\email{jhyu@itp.ac.cn}
\affiliation{
CAS Key Laboratory of Theoretical Physics, Institute of Theoretical Physics, \\ Chinese Academy of Sciences, Beijing 100190, P.\ R.\ China}
\affiliation{
School of Physical Sciences, University of Chinese Academy of Sciences, Beijing 100049, P.\ R.\ China}
\affiliation{
School of Fundamental Physics and Mathematical Sciences, Hangzhou Institute for Advanced
Study, UCAS, Hangzhou 310024, China}

\begin{abstract}

We identify an extended Poincare symmetry $ISO(2) \times ISO(3,1)$ for on-shell massive scattering amplitudes, transforming under the $U(2)$ Little group symmetry. Thus the particle state involves in both spin and transversality $t$, and the spin-spinors are extended to the spin-transverality spinors. The massive spin-$s$ spinors with different transversality can be related by the $SO(5,1)$ symmetry, although the reduced $U(2)$ Little group breaks the symmetry explicitly. The three-point massive amplitudes can be fully determined from the $T^\pm$ and $m$ generators, diagrammatically, denoted as the mass insertion and chirality flip, which provide correspondence between massless ultraviolet and massive chiral-eigenstate amplitudes. Thus the massless on-shell technique can be utilized to construct higher-point tree- and loop-level massive amplitudes.

\end{abstract}


\maketitle


\textbf{\textit{  Introduction.}} Modern S-matrix program explores the underling structure of scattering amplitudes consistent with unitarity, locality, causality and space-time symmetry. Both Feynman diagram and on-shell techniques have been utilized to calculate scattering amplitudes. The Feynman diagram method makes the locality and unitarity manifest, but not all the symmetries manifest, sometimes resulting extremely inefficient calculations. On the other hand, although the on-shell method does not have manifest locality due to spurious poles in amplitudes, but all the symmetries manifest, which leads to very efficient calculation~\cite{Parke:1986gb,Bern:1996je,Dixon:1996wi,Elvang:2013cua,Cheung:2017pzi,Travaglini:2022uwo}. For massless particles, the three-point (3-pt) amplitudes are entirely fixed by the consistence condition from little group scaling~\cite{Benincasa:2007xk,Witten:2003nn}, which exhibits the power of Poincare symmetry, while the four- and higher-particle amplitudes can be determined from factorization imposed by locality and unitarity through gluing technique or recursive relation~\cite{Britto:2004ap,Britto:2005fq}.

Given the fact that massless 3-pt amplitudes can be determined from Poincare symmetry, it is quite natural and tempting to extend this to massive amplitudes. There were earlier attempts~\cite{Kleiss:1985yh,Hagiwara:1985yu,Kleiss:1988xr,Dittmaier:1998nn,Schwinn:2005pi,Schwinn:2006ca,Badger:2005zh,Badger:2005jv,Conde:2016vxs} in which the $SU(2)$ Little group (LG) covariance was not manifest until the introduction of spin-spinors by Arkani-Hamed, Huang and Huang (AHH)~\cite{Arkani-Hamed:2017jhn}. In the spin-spinor formalism, the Lorentz structures of massive 3-pt amplitudes have been enumerated, although some redundancies needs to be removed~\footnote{Actually the AHH formalism usually gives over-complete 3-pt amplitudes, and the equation of motion relation should be applied to remove such redundancy, see Ref.~\cite{Durieux:2020gip} for examples. This indicates that the Poincare symmetry cannot fully determine the 3-pt amplitudes. }. The higher point amplitudes can be calculated, however, when massless internal photon is involved such as the simplest $e^+e^- \to \mu^+\mu^-$ process, challenges have been encountered since it cannot reproduce the Feynman diagram results~\cite{Christensen:2022nja,Lai:2023upa,Ema:2024vww}.

There are many physical processes involving in the chirality flip effect with mass enhancement, such as the charged pion decay, muon magnetic moment~\cite{Stockinger:2022ata, Arkani-Hamed:2021xlp}, flavor physics~\cite{Hall:1985dx,Gabbiani:1996hi,Misiak:1997ei}, etc. This indicates the massive amplitudes contain not only Lorentz structures as the massless ones, but also the chirality flip info, which is missing in the AHH formalism~\cite{Arkani-Hamed:2017jhn}. The chirality flip is manifest in the 'chiral-eigenstates' basis~\footnote{This chiral eigenstate basis has been utilized as the mass insertion~\cite{Dreiner:2008tw} in the Feynman diagram method. The 'chiral-eigenstates' basis is also called the 'gauge eigenstates' or 'interaction' basis in some literature~\cite{Hall:1985dx,Gabbiani:1996hi,Misiak:1997ei}, which is taken to be before mass diagnalization, and thus the off-diagonal part is from the Higgs vacuum expectation value (VEV) or flavor mixing if multi-flavor is taken. In this work, for simplicity, the multi-flavor case is not considered. }, while it is hidden in the 'mass-eigenstates' basis due to chirality mixing by equation of motion (EOM) relation, and Higgs VEV in symmetry breaking case.


In this work we recognize a new internal $U(1)_{D_-}$ symmetry with new quantum number {\it transversality} $t$ for massive scattering amplitudes, after lifting real mass parameters to complex mass spurions $(m, \tilde{m})$~\footnote{Taking the masses real or complex in the massive amplitudes should give equivalent infrared Lorentz structures. However, recognizing the symmetry for the complex masses plays the key role of identifying the chirality flip info, such as the mass enhancement, for the massive amplitudes. }. The $D_-$ charge on the $(m, \tilde{m})$ is similar to supersymmetric R-charge on $(Q, \bar{Q})$ space, and thus the momenta space $p_\mu$ is extended to $P_M = (p_\mu, m, \tilde{m})$, indicating an $ISO(5,1)$ space-time symmetry. This symmetry suggests the spin-spinors $\lambda_\alpha^I$ and $\tilde{\lambda}_{\dot{\alpha}}^I$ in the same multiplet~\footnote{Here the spin-spinors $\lambda_\alpha^I$ and $\tilde{\lambda}_{\dot{\alpha}}^I$ are extended to {\it the spin-chirality spinors} and further {\it the spin-transversality spinors}. Transversality and chirality are related, but {\it with differences}: $\lambda_\alpha^I$ and $\tilde{\lambda}_{\dot{\alpha}}^I$ carry both transversality and chirality, while $m$ and $\tilde{m}$ do not carry chirality, but carry transversality. The spinors with different transversality are related by a residue symmetry of the $SO(5,1)$ symmetry, isomorphism to 4-dimensional (4d) conformal symmetry.}. Unlike the AHH formalism only enumerates various possible 3-pt amplitudes, {\it this $SO(5,1)$ symmetry completely determines the Lorentz structures} of the 3-pt massive amplitudes  by the $T^\pm$ generators.

The 3-pt massive amplitudes should be extended to describe the chirality flip effects. This can be realized by the residue symmetry of the $ISO(5,1)$. Since keeping the LG keeping $p_M$ invariant is not $SU(2) \times SU(2)$, instead, a diagonal subgroup $SU(2)$, the $ISO(5,1)$ symmetry should be broken down to the $ISO(2) \times ISO(3,1)$, which avoids the Coleman-Mandula theorem~\cite{Coleman:1967ad}. This residue symmetry completely determines the 3-pt massive amplitudes with chirality flip. Furthermore, it could be applied to higher point tree and loop amplitudes with chirality flip enhancements.

In the high energy limit, the chirality and helicity should be unified, and thus lead to a one-to-one correspondence between massless and massive amplitudes, while in the AHH formalism, one massive amplitude could correspond to many kinds of ultraviolet (UV) massless ones, denoted as the infrared (IR) unification. The $e^+e^- \to \mu^+\mu^-$ does not match to the correct result~\cite{Christensen:2022nja} since the massive $FF\gamma$ amplitudes contains unwanted UV massless ones in the AHH formalism. Instead chirality help eliminate all irrelevant UV amplitudes and thus gives correct QED results.


\textbf{\textit{  Extended Poincar\'e Symmetry.}} In the 4d spinor-helicity formalism, a massive momentum $p$ can be decomposed as direct product of two massive spin-spinors $p_{\alpha\dot{\alpha}} \equiv \lambda_{\alpha}^I \tilde{\lambda}_{\dot{\alpha} I}$. The spin-spinors $\lambda_{\alpha}^{I}$ transform under the complex $SL(2,\mathbbm{C})$ spinor space and the $SU(2)$ LG keeping $p_{\alpha\dot{\alpha}}$ invariant:
\bea
\lambda_{\alpha}^I(p)\rightarrow \Lambda_{\alpha}^{\beta}\lambda_{\beta}^I(p)=\lambda_{\alpha}^J(\Lambda p)W(\Lambda;p)_J^I\;, 
\label{eq:spinor-transform}
\eea
similarly for $\tilde{\lambda}_{\dot{\alpha} I}$. The  Lorentz generators can be expressed by the spin-spinor variables as
\begin{align}
    L_{\alpha \beta}&=-i\left(\lambda_{\alpha}^{I}\frac{\partial}{\partial \lambda^{\beta I}}+\lambda_{\beta}^{I}\frac{\partial}{\partial \lambda^{\alpha I}}\right), 
\end{align}
and similarly ${\widetilde{L}}_{\dot{\alpha}\dot{\beta}}$ with $\lambda_{\alpha}^{I}$ replaced by $\tilde{\lambda}_{\dot{\alpha}}^{I}$.
The on-shell mass condition tells $p^2 =  \operatorname{det} \lambda \times \operatorname{det} \tilde{\lambda} = m \times \tilde{m} = m^2$, where $m=\frac{1}{2}\lambda_{\alpha I}\lambda^{\alpha I}$ and $\tilde{m}=\frac{1}{2} \tilde{\lambda}_{\dot{\alpha}I}\tilde{\lambda}^{\dot{\alpha}I}$ are two Lorentz invariants, with $m = \tilde{m} = \mathbf{m}$ for real momenta~\footnote{For real momenta $p_{\alpha \dot{\alpha}}$ is Hermitian, which imply $\tilde{\lambda}^I_{\dot{\alpha}}= \pm\left(\lambda^{I*}\right)_{\dot{\alpha}}$}.

In this work we propose the mass spurions $m$ and $\tilde{m}$ can be extended to be complex. Thus $m$ and $\tilde{m}$ carry equal but opposite phase $\tilde{m} = m^*$, and a $U(1)$ symmetry is identified: $m\rightarrow e^{-i\phi}m\;,  \tilde{m}\rightarrow e^{i\phi}\tilde{m}$, and correspondingly, the spin-spinors would be extended to have additional $U(1)$ transformation
\bea
\lambda_{\alpha}^{I}\rightarrow e^{-\frac{i}{2}\phi}\lambda_{\alpha}^{I}\;, \quad \tilde{\lambda}_{\dot{\alpha}I}\rightarrow e^{\frac{i}{2}\phi}\tilde{\lambda}_{\dot{\alpha}I}.
\label{eq:U1-transform}
\eea
Taking $\lambda_{\alpha}^I$ and $\tilde{\lambda}_{\dot{\alpha} I}$ are eigenstates of the $U(1)$ generator, we obtain
\bea
    D_-\equiv \frac{1}{2}\left(-\lambda_{\alpha}^{I}\frac{\partial}{\partial \lambda_{\alpha}^{I} }+\tilde{\lambda}_{\dot{\alpha}}^{I}\frac{\partial}{\partial \tilde{\lambda}_{\dot{\alpha}}^{I}}\right),
\eea
which commutates with the Lorentz generators $L_{\alpha \beta}$ and ${\widetilde{L}}_{\dot{\alpha}\dot{\beta}}$, and thus the extended symmetry is $SO(2) \times SO(3,1)$, in which isomorphism
$SO(2) \cong U(1)$ and $SO(3,1) \cong SL(2, \mathbb{C})$ are taken.
This $U(1)$ symmetry identify a new quantum number {\it transversality} $t$ with $t=\mp \frac12$ for $\lambda_{\alpha}^I (\tilde{\lambda}_{\dot{\alpha} I})$. 
Since $e^{i\phi D_-}me^{-i\phi D_-}=me^{-i\phi}$, the $U(1)$ generator $D_-$, together with $m$ and $\tilde{m}$, composes the Lie algebra of a $ISO(2)$ group:
\bea
    [D_-,m]=-m\;,\quad [D_-,\tilde{m}]=+\tilde{m}\;,\quad [m,\tilde{m}]=0.
\eea
This is similar to the $ISO(2)$ symmetry in the supersymmetric $Q$ and $\bar{Q}$ with the $U(1)_R$ charge. Therefore, like the superspace, the momentum $p_\mu$ can be extended as $P_M = (p_\mu, m, \tilde{m})$.

Since $\lambda_{\alpha}^I$ and $\tilde{\lambda}_{\dot{\alpha} I}$ carry opposite transversality $t$ charge, it motivates us to relate these two spinors into one multiplet, such that $\lambda^{I}_A=\{\lambda_{\alpha}^{I}, \tilde{\lambda}^{\dot{\alpha}I}\}$, with index $A\equiv\{\alpha, \dot{\alpha}\}$. To realize such relation we consider two new generators $T^\pm$ 
\bea
    T^{+}_{\alpha \dot{\alpha}}\equiv \tilde{\lambda}_{\dot{\alpha}}^{I}\frac{\partial}{\partial \lambda^{\alpha I}},\quad
    T^{-}_{\alpha \dot{\alpha}}\equiv \lambda_{\alpha}^{I}\frac{\partial}{\partial \tilde{\lambda}^{\dot{\alpha}I}},
\eea
which would raise/lower the transversality due to the commutator relation $[D_-,T^\pm]=\pm T^\pm$. From the commutator relation
\begin{equation}
    [(T^+)_{\alpha\dot{\alpha}}, (T^-)_{\beta\dot{\beta}} ] = -\frac{i}{2}( \epsilon_{\alpha\beta} \tilde{L}_{\dot{\alpha}\dot{\beta}} +\tilde{\epsilon}_{\dot{\alpha}\dot{\beta}} L_{\alpha\beta} ) -\frac{1}{2}\epsilon_{\alpha\beta}\tilde{\epsilon}_{\dot{\alpha}\dot{\beta}} D_-,
\end{equation}
we find the $D_-, T^\pm, L$ generators form $SO(5,1)$ Lie algebra.

Let us consider the irreducible representations (irreps) of the $SO(5,1)$ symmetry. Since it is isomorphism to the 4d conformal symmetry, we would like to use three quantum numbers $(\Delta,J_1,J_2)$ to label irreps, in which $\Delta$ labels the maximal transversality, and $(J_1,J_2)$ labels the Lorentz irreps. The primary irrep should be $\Delta = s$ while there are infinite descendant irreps increasing $\Delta$. Each irrep $(\Delta,J_1,J_2)$ could be decomposed as the irreps $[t,j_1,j_2]$ of $SO(2)\times SO(3,1)$. The $(\Delta,0,\frac{1}{2})$ irrep with $\Delta=\frac{1}{2},\frac{3}{2}$ for spin-$1/2$ particle has
\bea
\begin{tikzpicture}[baseline=-0.5cm,scale=0.8,transform shape]
\path 
(0,0) node(A1) [rectangle] {$[\frac12,0,\frac12]$}
(2.3,0) node(A2) [rectangle] {$[-\frac12,\frac12,0]$}
(-2.1,-1.05) node(B1) [rectangle] {$[\frac32,0,\frac12]$}
(0,-0.8) node(B2) [rectangle] {$[\frac12,\frac12,1]$}
(0,-1.3) node(B3) [rectangle] {$[\frac12,\frac12,0]$}
(2.3,-0.8) node(B4) [rectangle] {$[-\frac12,1,\frac12]$}
(2.3,-1.3) node(B5) [rectangle] {$[-\frac12,0,\frac12]$}
(4.6,-1.05) node(B6) [rectangle] {$[-\frac32,\frac12,0]$};
\draw[color=red,->] (A1)--node [color=black,above]{$T^-$} (A2);
\draw[color=red,->] (B1)--node [color=black,above]{$T^-$} (-0.6,-1.05);
\draw[color=red,->] (0.7,-1.05)--node [color=black,above]{$T^-$} (1.5,-1.05);
\draw[color=red,->] (3.1,-1.05)--node [color=black,above]{$T^-$} (B6);
\end{tikzpicture}
\eea
Similarly the $(\Delta,0,1)$ irrep with $\Delta=1,2,3$ for the spin-$1$:
\bea
\begin{tikzpicture}[baseline=-1.1cm,scale=0.6,transform shape]
\path 
(0,0) node(A1) [rectangle] {$[1,0,1]$}
(2,0) node(A2) [rectangle] {$[0,\frac12,\frac12]$}
(4,0) node(A3) [rectangle] {$[-1,1,0]$}
(-2,-1.3) node(B1) [rectangle] {$[2,0,1]$}
(0,-1.05) node(B2) [rectangle] {$[1,\frac12,\frac12]$}
(0,-1.55) node(B3) [rectangle] {$[1,\frac32,\frac12]$}
(2,-0.8) node(B4) [rectangle] {$[0,1,0]$}
(2,-1.3) node(B5) [rectangle] {$[0,1,1]$}
(2,-1.8) node(B6) [rectangle] {$[0,0,1]$}
(4,-1.05) node(B7) [rectangle] {$[-1,\frac12,\frac12]$}
(4,-1.55) node(B8) [rectangle] {$[-1,\frac32,\frac12]$}
(6,-1.3) node(B9) [rectangle] {$[-2,1,0]$}
(-4,-3.3) node(C1) [rectangle] {$[3,0,1]$}
(-2,-3.05) node(C2) [rectangle] {$[2,\frac12,\frac32]$}
(-2,-3.55) node(C3) [rectangle] {$[2,\frac12,\frac12]$}
(0,-2.55) node(C4) [rectangle] {$[1,1,0]$}
(0,-3.05) node(C5) [rectangle] {$[1,1,1]$}
(0,-3.55) node(C6) [rectangle] {$[1,0,1]$}
(0,-4.05) node(C7) [rectangle] {$[1,1,2]$}
(2,-2.55) node(C8) [rectangle] {$[0,\frac32,\frac12]$}
(2,-3.05) node(C9) [rectangle] {$[0,\frac32,\frac32]$}
(2,-3.55) node(C10) [rectangle] {$[0,\frac12,\frac12]$}
(2,-4.05) node(C11) [rectangle] {$[0,1,1]$}
(4,-2.55) node(C12) [rectangle] {$[-1,1,0]$}
(4,-3.05) node(C13) [rectangle] {$[-1,0,1]$}
(4,-3.55) node(C14) [rectangle] {$[-1,1,1]$}
(4,-4.05) node(C15) [rectangle] {$[-1,2,1]$}
(6,-3.05) node(C16) [rectangle] {$[-2,\frac32,\frac12]$}
(6,-3.55) node(C17) [rectangle] {$[-2,\frac12,\frac12]$}
(8,-3.3) node(C18) [rectangle] {$[-3,0,1]$};
\draw[color=red,->] (A1)--node [color=black,above]{$T^-$} (A2);
\draw[color=red,->] (A2)--node [color=black,above]{$T^-$} (A3);
\draw[color=red,->] (B1)--node [color=black,above]{$T^-$} (-0.6,-1.3);
\draw[color=red,->] (0.7,-1.3)--node [color=black,above]{$T^-$} (B5);
\draw[color=red,->] (B5)--node [color=black,above]{$T^-$} (3.3,-1.3);
\draw[color=red,->] (4.7,-1.3)--node [color=black,above]{$T^-$} (5.3,-1.3);
\draw[color=red,->] (C1)--node [color=black,above]{$T^-$} (-2.7,-3.3);
\draw[color=red,->] (-1.3,-3.3)--node [color=black,above]{$T^-$} (-0.6,-3.3);
\draw[color=red,->] (0.6,-3.3)--node [color=black,above]{$T^-$} (1.3,-3.3);
\draw[color=red,->] (2.6,-3.3)--node [color=black,above]{$T^-$} (3.3,-3.3);
\draw[color=red,->] (4.7,-3.3)--node [color=black,above]{$T^-$} (5.3,-3.3);
\draw[color=red,->] (6.7,-3.3)--node [color=black,above]{$T^-$} (7.3,-3.3);
\end{tikzpicture}
\eea


\textbf{\textit{  Spin-Transversality Spinors.}} The $\lambda^{I}_A=\{\lambda_{\alpha}^{I}, \tilde{\lambda}^{\dot{\alpha}I}\}$ transforms under the fundamental (fund) rep of the $SO(5,1)$ spinor space and the $SU(2)$ LG space. If we introduce momenta $p_{AB}$, since it can be decomposed as direct product of fund and anti-fund reps, we should also have inequivalent $\overline{\lambda}^{\overline{I}}_B$ transforming under another $SU(\overline{2})$ Little group space, which constitutes $SU(2) \times SU(\overline{2})$ LG for $ISO(5,1)$~\footnote{In fact, we should distinguish different LG space. Any momentum $p^{AB}$ can be decomposed as $p^{AB} = \chi^{A I_l} \chi^B_{I_l}$ and $p_{AB} = \tilde{\chi}_A^{I_r} \tilde{\chi}_{BI_r}$, where $\chi^{A I_l}$ and $\tilde{\chi}_A^{I_r}$ are inequivalent fund and anti-fund reps. Note the spin indices $I_l$ and $I_r$ are different from the index $I$ in Eq.~(\ref{eq:spinor-transform}) since the LG $SU(2)$ is the diagonal subgroup of the $SU(2)_l \times SU(2)_r$ one.}. 

According to Wigner's construction~\cite{Wigner:1939cj}, a physical state should be described by induced rep of space-time symmetry. Based on Eq.~\eqref{eq:spinor-transform} and \eqref{eq:U1-transform}, the physical state should be $|p, t, s, s_z \rangle$, described by the $U(2)$ LG keeping $p_\mu$ invariant. Thus the reduced $U(2)$ Little group indicates the $ISO(5,1)$ symmetry broken down to $SO(2) \times ISO(3,1)$~\footnote{This symmetry breaking can also be seen from the contradiction between the $U(2)$ LG and the $SU(2) \times SU(\overline{2})$ LG which is the induced rep of the $ISO(5,1)$. In fact, the symmetry breaking induces $SU(2)_l \times SU(2)_r \to SU(2)$. }. There is an equivalent description on physical states. Taking the maximized symmetry $ISO(2) \times ISO(3,1)$, the physical state should be described by $|p, m, \tilde{m}, s, s_z \rangle$ which belongs to the $SU(2)$ LG keeping $p_M$ invariant. In this way, the transversality transformation is under the $SO(2) \times SO(3,1)$ spinor space, not the LG space. This motivates us to connect transversality to chirality, which characterize the spinor irreps.

In QFT, the spinors $\lambda^I$ and $\tilde{\lambda}^I$ are spin-$\frac12$ wave-functions for fermion with left-handed $L$ and right-handed $R$ chirality, or anti-fermion with opposite chirality. On the other hand, we take $\lambda^I$ and $\tilde{\lambda}^I$ as the $[t, j_1, j_2]$ irreps 
\bea
{\textrm{spin-}}1/2: && \quad \tilde{\lambda}^I \ \ [1/2, 0, 1/2], \ \ \lambda^I \ \  [-1/2, 1/2, 0], \label{eq:primaryhalf-transversality}
\eea
Thus transversality is closely related to chirality: the $L$ ($R$) chirality corresponds to $t= -\frac12$ ($+\frac12$) for fermion, verse vice for anti-fermion. Generalizing the chirality from spin-$1/2$ to spin-$s$ particle, and identify chirality as $t = j_2- j_1$ for fermion, we introduce {\it the spin-chirality spinor} in $[t, j_1, j_2]$ rep~\footnote{Note the spinor carrying chirality and transversality have the same form as AHH spinor, because the $U(1)$ transversality phase can be absorbed into spinor due to complex nature of $\lambda^{I}$ and $\tilde{\lambda}^{I}$. It transforms under both the irrep $[t, j_1, j_2]$ of the $SO(2) \times SO(3,1)$ spinor space, and the irreps $[s, \sigma]$ of the $SU(2)$ little group. }
\bea
\lambda_{\alpha_1}^{(I_1}\dots \lambda_{\alpha_{2j_1}}^{I_{2j_1}}\Tilde{\lambda}_{\dot{\alpha}_1}^{I_{2j_1+1}}\dots\Tilde{\lambda}_{\dot{\alpha}_{2j_2}}^{I_{2s)}},
\eea
with symmetrized indices taken. For spin-$s$ particle, there are $(2s+1)$ spin-chirality spinors, related by $T^\pm$ generators.

Note for each $(j_1, j_2)$, the spin-chirality spinors only describe one kind of transversality, and there should have $2s+1$ spinors running over all the transversality. So the spin-chirality spinor should be extended to {\it the spin-transversality (ST) spinor}, in which not only the primary, but also the descendant irreps should be taken, with {\it the selection condition}: 
\bea
s = j_1 + j_2, \  |t|\leq s, \  s\leq\Delta\leq 3s, \  |t-j_2+j_1|=\Delta-s. 
\label{eq:so51condition}
\eea
Note $m$ and $\tilde{m}$ do not carry chirality, so the selected descendant spinors belongs to the same spin-chirality spinors, but different spin-transversality spinors. For spin-$1/2$ particle, we contract $\lambda^{I}$ and $\tilde{\lambda}^{I}$ with $p_{\alpha \dot{\alpha}}$~\footnote{Here the equation of motion
$p_{\alpha \dot{\alpha}} \tilde{\lambda}^{\dot{\alpha} I}= \tilde{m} \lambda_{\alpha}^{I}, \ p_{\alpha \dot{\alpha}} \lambda^{\alpha I}=-m \tilde{\lambda}_{\dot{\alpha}}^{I}$
is utilized}
to obtain other transversality
\bea
{\textrm{spin-}}1/2: && \quad \tilde{m}\lambda^I \ [1/2, 1/2, 0], \ \ m \tilde{\lambda}^I \ [-1/2,  0, 1/2].  \label{eq:descendanthalf-transversality}
\eea
Together with Eq.~\eqref{eq:primaryhalf-transversality}, we complete the ST spinors for spin-$1/2$ particle. This provides us a general way of writing the ST spinors: start from primary rep with highest $t$, and apply $T^-$ to obtain all primary reps, then contract indices with $p$ to obtain descendant reps, and select the ones satisfying the condition in Eq.~\eqref{eq:so51condition}. For spin-$1$, the complete ST spinors are
\bea
\resizebox{.8\linewidth}{!}{
\begin{tabular}{c|c|c|c}
\hline
& $t = 1$ & $t = 0$ & $t = -1$ \\
\hline
$\Delta = 1$ & $\tilde{\lambda}^{(I_{1}}_{\dot{\alpha}_{1}}\tilde{\lambda}^{I_{2})}_{\dot{\alpha}_{2}}$ & $\lambda^{(I_{1}}_{\alpha_{1}}\tilde{\lambda}^{I_{2})}_{\dot{\alpha}_{1}}$ & $\lambda^{(I_{1}}_{\alpha_{1}}\lambda^{I_{2})}_{\alpha_{2}}$ \\
$\Delta= 2$ & $\lambda^{(I_{1}}_{\alpha_{1}}\tilde{\lambda}^{I_{2})}_{\dot{\alpha}_{1}}\tilde{m}$  & $\tilde{\lambda}^{(I_{1}}_{\dot{\alpha}_{1}}\tilde{\lambda}^{I_{2})}_{\dot{\alpha}_{2}}m,\lambda^{(I_{1}}_{\alpha_{1}}\lambda^{I_{2})}_{\alpha_{2}}\tilde{m}$ & $\lambda^{(I_{1}}_{\alpha_{1}}\tilde{\lambda}^{I_{2})}_{\dot{\alpha}_{1}}m$ \\
$\Delta = 3$ & $\lambda^{(I_{1}}_{\alpha_{1}}\lambda^{I_{2})}_{\alpha_{2}}\tilde{m}^2$  &  & $\tilde{\lambda}^{(I_{1}}_{\dot{\alpha}_{1}}\tilde{\lambda}^{I_{2})}_{\dot{\alpha}_{2}}m^2$ \\
\hline
\end{tabular}}
\eea

There are differences among the AHH spinor, ST spinor,  and $SO(5,1)$ spinor: AHH spinor only takes one $[t, j_1, j_2]$ irrep (usually the $[-t_{\max},{j_1}_{\max},0]$ irrep), other irrep can be freely converted into this irrep by EOM; ST spinors need all $(2s+1)^2$ irreps; $SO(5,1)$ spinors contain complete $(\Delta, J_1, J_2)$ primary and descendant reps without the selection condition.  

\textbf{\textit{  On-shell Mass Insertion and Chirality Flip.}} The above spin-transversality spinors can be presented diagrammatically. Let us first consider the external on-shell particle. 
Diagrammatically, spin-$\frac12$  wave-functions $\lambda$ and $\tilde{\lambda}$ take
\begin{equation} \begin{aligned} \label{eq:fermion}
\begin{tabular}{l|l|l}
$t$ & Fermion & Antifermion \\
\hline
$-\frac{1}{2}$ & 
L: \begin{tikzpicture}[baseline=0.7cm] \begin{feynhand}
\vertex [particle] (i1) at (1.35,0.8) {};
\vertex [dot] (v1) at (0,0.8) {};
\graph{(v1)--[plain,red,very thick] (i1)};
\end{feynhand} \end{tikzpicture}$=\lambda^I_\alpha$ & 
R: \begin{tikzpicture}[baseline=0.7cm] \begin{feynhand}
\vertex [particle] (i1) at (1.35,0.8) {};
\vertex [dot] (v1) at (0,0.8) {};
\graph{(v1)--[plain,cyan,very thick] (i1)};
\end{feynhand} \end{tikzpicture}$=\lambda^{I\alpha}$ \\
$+\frac{1}{2}$ &
R: \begin{tikzpicture}[baseline=0.7cm] \begin{feynhand}
\vertex [particle] (i1) at (1.35,0.8) {};
\vertex [dot] (v1) at (0,0.8) {};
\graph{(v1)--[plain,cyan,very thick] (i1)};
\end{feynhand} \end{tikzpicture}$=\tilde{\lambda}^I_{\dot{\alpha}}$ & 
L: \begin{tikzpicture}[baseline=0.7cm] \begin{feynhand}
\vertex [particle] (i1) at (1.35,0.8) {};
\vertex [dot] (v1) at (0,0.8) {};
\graph{(v1)--[plain,red,very thick] (i1)};
\end{feynhand} \end{tikzpicture}$=\tilde{\lambda}^{I\dot{\alpha}}$ \\
\end{tabular}
\end{aligned} \end{equation}

After drawing wave-functions for external (anti-)fermions, let us construct the internal fermion line by gluing the fermion and antifermion with the same (and opposite) chirality. The inner product of the same chirality particles should have the same Lorentz rep $(j_1, j_2)$ with summing over spin indices
\begin{equation} 
\label{eq:onshellpropogator-momentum}
\begin{aligned}
\begin{tikzpicture}[baseline=0.7cm] \begin{feynhand}
\vertex [dot] (v1) at (0,0.8) {};
\vertex [dot] (v2) at (1.4,0.8) {};
\graph{(v2)--[plain,red,very thick](v1)};
\end{feynhand} \end{tikzpicture}&=\lambda^I_\alpha \tilde{\lambda}^{\dot{\beta}}_I=p_{\alpha}^{\dot{\beta}},\\
\begin{tikzpicture}[baseline=0.7cm] \begin{feynhand}
\vertex [dot] (v1) at (0,0.8) {};
\vertex [dot] (v2) at (1.4,0.8) {};
\graph{(v2)--[plain,cyan,very thick] (v1)};
\end{feynhand} \end{tikzpicture}&=\tilde{\lambda}_{\dot{\alpha}}^I \lambda_I^\beta =-p_{\dot{\alpha}}^{\beta}.
\end{aligned} \end{equation}
On the other hand, the inner product of the opposite chirality would obtain {\it the mass insertion}:
\begin{equation} 
\label{eq:onshellpropogator-mass}
\begin{aligned}
\begin{tikzpicture}[baseline=0.7cm] \begin{feynhand}
\vertex [dot] (i1) at (1.4,0.8) {};
\vertex [dot] (v1) at (0,0.8) {};
\vertex (v2) at (0.7,0.8);
\graph{(v1)--[plain,cyan,very thick] (v2)--[plain,red,very thick] (i1)};
\draw[very thick] plot[mark=x,mark size=2.7] coordinates {(v2)};
\end{feynhand} \end{tikzpicture}
&=\tilde{\lambda}_{\dot{\alpha}}^I \tilde{\lambda}^{\dot{\beta}}_I=\tilde{m} \delta_{\dot{\alpha}}^{\dot{\beta}},\\
\begin{tikzpicture}[baseline=0.7cm] \begin{feynhand}
\vertex [dot] (i1) at (1.4,0.8) {};
\vertex [dot] (v1) at (0,0.8) {};
\vertex (v2) at (0.7,0.8);
\graph{(v1)--[plain,red,very thick] (v2)--[plain,cyan,very thick] (i1)};
\draw[very thick] plot[mark=x,mark size=2.7] coordinates {(v2)};
\end{feynhand} \end{tikzpicture}
&=\lambda^I_\alpha \lambda_I^\beta=m \delta_{\alpha}^\beta,
\end{aligned} \end{equation}
where the chirality is flipped by the mass term.

For descendant ones $\tilde{m}\lambda^I$ and $m \tilde{\lambda}^I$, consider external fermion with $R$ chirality (cyan), the chirality is flipped to internal red line via the inner product between internal $p$ and external $\tilde{\lambda}$. It is equivalent to the inner product of internal $L$ fermion and mass insertion denoting chirality flip: 
\begin{equation}
\begin{tikzpicture}[baseline=0.7cm] \begin{feynhand}
\vertex [particle] (i1) at (1.55,0.8) {};
\vertex [dot] (v1) at (0,0.8) {};
\vertex (v2) at (0.75,0.8);
\graph{(v1)--[plain,red,very thick] (v2)--[plain,cyan,insertion={[style=black]0},very thick] (i1)};
\draw[snake=brace] (0.1,0.9) -- (0.6,0.9);
\draw[snake=brace] (0.9,0.9) -- (1.4,0.9);
\path (0.35,1.25) node(A0) [rectangle] {$p^{\dot{\alpha}}_\alpha$}
(1.15,1.25) node(A1) [rectangle] {$\tilde{\lambda}^I_{\dot{\alpha}}$};
\end{feynhand} \end{tikzpicture} \xrightarrow{\textrm{EOM}} 
\begin{tikzpicture}[baseline=0.7cm] \begin{feynhand}
\vertex [particle] (i1) at (1.55,0.8) {};
\vertex [dot] (v1) at (0,0.8) {};
\vertex (v2) at (1,0.8);
\graph{(v1)--[plain,red,very thick] (v2)--[plain,cyan,insertion={[style=black]0},very thick] (i1)};
\draw[snake=brace] (0.1,0.9) -- (0.5,0.9);
\draw[snake=brace] (0.6,0.9) -- (1.4,0.9);
\path (0.35,1.25) node(A0) [rectangle] {$\lambda^I_\alpha$}
(1,1.25) node(A1) [rectangle] {$\tilde{m}$};
\end{feynhand} \end{tikzpicture}
\end{equation}
Using this notation, we draw diagrams for the complete spin-transversality spinors for spin-$\frac12$ particle
\bea \label{eq:Fdiagram}
\begin{tikzpicture}[baseline=0.7cm] \begin{feynhand}
\vertex [particle] (i1) at (1.35,1.6) {};
\vertex [dot] (v1) at (0,1.6) {};
\graph{(v1)--[plain,cyan,very thick] (i1)};
\vertex [particle] (i2) at (4.35,1.6) {};
\vertex [dot] (v2) at (3,1.6) {};
\graph{(v2)--[plain,red,very thick] (i2)};
\vertex [particle] (i3) at (1.35,0) {};
\vertex [dot] (v3) at (0,0) {};
\vertex (j3) at (0.6,0);
\graph{(v3)--[plain,red,very thick] (j3)--[plain,cyan,very thick] (i3)};
\draw[very thick] plot[mark=x,mark size=2.7] coordinates {(j3)};
\vertex [particle] (i4) at (4.35,0) {};
\vertex [dot] (v4) at (3,0) {};
\vertex (j4) at (3.6,0);
\graph{(v4)--[plain,cyan,very thick] (j4)--[plain,red,very thick] (i4)};
\draw[very thick] plot[mark=x,mark size=2.7] coordinates {(j4)};
\path 
(0.6,1.34) node(A1) [rectangle,scale=0.8] {$\tilde{\lambda}^I_{\dot{\alpha}}$}
(3.6,1.34) node(A2) [rectangle,scale=0.8] {$\lambda^I_{\alpha}$}
(0.6,-0.32) node(A3) [rectangle,scale=0.8] {$\tilde{m}\lambda^I_{\alpha}$}
(3.6,-0.32) node(A4) [rectangle,scale=0.8] {$m\tilde{\lambda}^I_{\dot{\alpha}}$}
(1.44,1.07) node(B1) [rectangle,scale=0.8] {$m$}
(2.68,1.1) node(B2) [rectangle,scale=0.8] {$\tilde{m}$};
\draw[-latex,dash pattern=on 1.5pt off 1pt,very thin] (1.35,1.6)--node[above,scale=0.8]{$T^-$}(2.8,1.6);
\draw[-latex,dash pattern=on 1.5pt off 1pt,very thin] (1.35,0)--node[above,scale=0.8]{$T^-$}(2.8,0);
\draw[-latex,dash pattern=on 1.5pt off 1pt,very thin] (1.25,1.5)--(2.9,0.1);
\draw[-latex,dash pattern=on 1.5pt off 1pt,very thin] (2.8,1.5)--(1.25,0.1);
\end{feynhand} \end{tikzpicture}
\eea
Similarly we obtain for spin-$1$ particle
\bea \label{eq:Vdiagram}
\begin{tikzpicture}[baseline=0cm] \begin{feynhand}
\vertex [particle] (i1) at (1.35,1.6) {};
\vertex [dot] (v1) at (0,1.6) {};
\graph{(v1)--[plain,cyan,very thick] (i1)};
\vertex [particle] (i2) at (4.35,1.6) {};
\vertex [dot] (v2) at (3,1.6) {};
\graph{(v2)--[plain,brown,very thick] (i2)};
\vertex [particle] (i3) at (7.35,1.6) {};
\vertex [dot] (v3) at (6,1.6) {};
\graph{(v3)--[plain,red,very thick] (i3)};
\vertex [particle] (i4) at (1.35,0) {};
\vertex [dot] (v4) at (0,0) {};
\vertex (j4) at (0.6,0);
\graph{(v4)--[plain,brown,very thick] (j4)--[plain,cyan,very thick] (i4)};
\draw[very thick] plot[mark=x,mark size=2.7] coordinates {(j4)};
\vertex [particle] (i5) at (4.35,0.2) {};
\vertex [dot] (v5) at (3,0.2) {};
\vertex (j5) at (3.6,0.2);
\graph{(v5)--[plain,cyan,very thick] (j5)--[plain,brown,very thick] (i5)};
\draw[very thick] plot[mark=x,mark size=2.7] coordinates {(j5)};
\vertex [particle] (i6) at (4.35,-0.2) {};
\vertex [dot] (v6) at (3,-0.2) {};
\vertex (j6) at (3.6,-0.2);
\graph{(v6)--[plain,red,very thick] (j6)--[plain,brown,very thick] (i6)};
\draw[very thick] plot[mark=x,mark size=2.7] coordinates {(j6)};
\vertex [particle] (i7) at (7.35,0) {};
\vertex [dot] (v7) at (6,0) {};
\vertex (j7) at (6.6,0);
\graph{(v7)--[plain,brown,very thick] (j7)--[plain,red,very thick] (i7)};
\draw[very thick] plot[mark=x,mark size=2.7] coordinates {(j7)};
\vertex [particle] (i8) at (1.6,-1.6) {};
\vertex [dot] (v8) at (0,-1.6) {};
\vertex (j8) at (0.5,-1.6);
\vertex (k8) at (1,-1.6);
\graph{(v8)--[plain,red,very thick] (j8)--[plain,brown,very thick] (k8)--[plain,cyan,very thick] (i8)};
\draw[very thick] plot[mark=x,mark size=2.7] coordinates {(j8)};
\draw[very thick] plot[mark=x,mark size=2.7] coordinates {(k8)};
\vertex [particle] (i9) at (7.35,-1.6) {};
\vertex [dot] (v9) at (5.75,-1.6) {};
\vertex (j9) at (6.25,-1.6);
\vertex (k9) at (6.75,-1.6);
\graph{(v9)--[plain,cyan,very thick] (j9)--[plain,brown,very thick] (k9)--[plain,red,very thick] (i9)};
\draw[very thick] plot[mark=x,mark size=2.7] coordinates {(j9)};
\draw[very thick] plot[mark=x,mark size=2.7] coordinates {(k9)};
\path 
(0.6,1.3) node(A1) [rectangle,scale=0.8] {$\tilde{\lambda}^{(I_1}_{\dot{\alpha}_1}\tilde{\lambda}^{I_2)}_{\dot{\alpha}_2}$}
(3.6,1.3) node(A2) [rectangle,scale=0.8] {$\tilde{\lambda}^{(I_1}_{\dot{\alpha}_1}\lambda^{I_2)}_{\alpha_2}$}
(6.6,1.3) node(A3) [rectangle,scale=0.8] {$\lambda^{(I_1}_{\alpha_1}\lambda^{I_2)}_{\alpha_2}$}
(0.6,-0.30) node(A4) [rectangle,scale=0.8] {$\tilde{m}\tilde{\lambda}^{(I_1}_{\dot{\alpha}_1}\lambda^{I_2)}_{\alpha_2}$}
(3.6,0.5) node(A5) [rectangle,scale=0.8] {$m\tilde{\lambda}^{(I_1}_{\dot{\alpha}_1}\tilde{\lambda}^{I_2)}_{\dot{\alpha}_2}$}
(3.6,-0.5) node(A6) [rectangle,scale=0.8] {$\tilde{m}\lambda^{(I_1}_{\alpha_1}\lambda^{I_2)}_{\alpha_2}$}
(6.6,-0.3) node(A7) [rectangle,scale=0.8] {$m\tilde{\lambda}^{(I_1}_{\dot{\alpha}_1}\lambda^{I_2)}_{\alpha_2}$}
(0.6,-1.9) node(A8) [rectangle,scale=0.8] {$\tilde{m}^2\lambda^{(I_1}_{\alpha_1}\lambda^{I_2)}_{\alpha_2}$}
(6.6,-1.9) node(A9) [rectangle,scale=0.8] {$m^2\tilde{\lambda}^{(I_1}_{\dot{\alpha}_1}\tilde{\lambda}^{I_2)}_{\dot{\alpha}_2}$}
(1.44,1.07) node(B1) [rectangle,scale=0.8] {$m$}
(2.7,1.1) node(B2) [rectangle,scale=0.8] {$\tilde{m}$}
(4.44,1.07) node(B3) [rectangle,scale=0.8] {$m$}
(5.7,1.1) node(B4) [rectangle,scale=0.8] {$\tilde{m}$}
(2.4,-1.0) node(B5) [rectangle,scale=0.8] {$\tilde{m}$}
(5.0,-1.0) node(B6) [rectangle,scale=0.8] {$m$};
\draw[-latex,dash pattern=on 1.5pt off 1pt,very thin] (1.35,1.6)--node[above,scale=0.8]{$T^-$}(2.8,1.6);
\draw[-latex,dash pattern=on 1.5pt off 1pt,very thin] (4.35,1.6)--node[above,scale=0.8]{$T^-$}(5.8,1.6);
\draw[-latex,dash pattern=on 1.5pt off 1pt,very thin] (1.35,0)--node[above,scale=0.8]{$T^-$}(2.8,0);
\draw[-latex,dash pattern=on 1.5pt off 1pt,very thin] (4.6,0)--node[above,scale=0.8]{$T^-$}(5.8,0);
\draw[-latex,dash pattern=on 1.5pt off 1pt,very thin] (1.6,-1.6)--node[below,scale=0.8]{$(T^-)^2$}(5.6,-1.6);
\draw[-latex,dash pattern=on 1.5pt off 1pt,very thin] (1.25,1.5)--(2.9,0.3);
\draw[-latex,dash pattern=on 1.5pt off 1pt,very thin] (2.8,1.5)--(1.25,0.1);
\draw[-latex,dash pattern=on 1.5pt off 1pt,very thin] (4.25,1.5)--(5.9,0.1);
\draw[-latex,dash pattern=on 1.5pt off 1pt,very thin] (5.8,1.5)--(4.27,-0.2);
\draw[-latex,dash pattern=on 1.5pt off 1pt,very thin] (2.85,-0.35)--(1.4,-1.5);
\draw[-latex,dash pattern=on 1.5pt off 1pt,very thin] (4.3,0.2)--(5.7,-1.45);
\end{feynhand} \end{tikzpicture}
\eea
where the $T^\pm$ generators relate different transversality for primary reps, while $m$ generator relate different transversality with the same $(j_1, j_2)$.

\textbf{\textit{  Massive 3-pt Amplitudes determined from Symmetry.}} Similar to the single particle spinor construction, the 3-pt massive amplitudes can be fully determined from the $T^\pm$ and $m$ generators.
The only difference is that the 3-pt amplitudes should be the Lorentz singlet and thus can be characterized only by the transversality $[t_1,t_2,t_3]$ with $-s_i\le t_i\le +s_i$, since for each particle $i$ once $t_i$ is fixed in each $\Delta$ the $(j_{1,i}, j_{2,i})$ is determined. Let us first determine irrep with maximal transversality, belonged to the primary irreps. Similar to 3-pt massless amplitudes determined from LG scaling, the $SU(2)$ LG transformation fully determines this amplitude~\footnote{First the 3-pt kinematics tells the following anti-holomorphic kinematic configuration ansatz
\begin{equation}
[s_1,s_2,s_3]: [\mathbf{12}]^{y_{12}}[\mathbf{23}]^{y_{23}}[\mathbf{31}]^{y_{31}}.
\end{equation} 
To determine the unknown $y_{ij}$, we consider the little group covariance for massive particles. In the LG of a spin-$s$ particle, the amplitude should transform as a rank-$2s$ tensors
\begin{equation}
2s_1=y_{12}+y_{31},\ \
2s_2=y_{12}+y_{23},\ \ 
2s_3=y_{23}+y_{31}.
\end{equation}
Therefore, the primary rep is determined by the little-group covariance.} 
\begin{equation}
[s_1,s_2,s_3]:\quad [\mathbf{12}]^{\Delta s_3}[\mathbf{23}]^{\Delta s_1}[\mathbf{31}]^{\Delta s_2},
\label{eq:3pt-maxtrans}
\end{equation}
where $\Delta s_i=s_1+s_2+s_3-s_i$~\footnote{For convenience, we set $s_1\ge s_2\ge s_3$ here. If the 3-point amplitude satisfies the condition $s_1\le s_2+s_3$, the internal structure can be always reduced to the trivial one, i.e. a product of Levi-Civita tensor. If this condition is not satisfied, we need to consider the non-trivial internal structure. We choose the internal structure to be the monomial in $\mathbf{p}_3$. In this case, we will begin with a different primary rep,
$[\mathbf{12}]^{2s_3} [\mathbf{23}]^{2s_2} [\mathbf{1}|\mathbf{p}_3|\mathbf{1}\rangle^{s_1-s_2-s_3}$. Then we can follow the same procedure to derive the complete amplitude.}. Here we use the \textbf{Bold} notation with $\langle\mathbf{12}\rangle = \langle 1^I 2^J\rangle$, so the little-group indices are dropped. In the primary irreps with Lorentz singlet, there are other massive amplitudes, which can be determined by the $T^-_i \cdot T^-_j$ generators. For 3-pt amplitudes, the generators should take $T^-_1 \cdot T^-_2$, $T^-_2 \cdot T^-_3$ and $T^-_3 \cdot T^-_1$. Acting these to Eq.~\eqref{eq:3pt-maxtrans}, we get other irreps with lower transversality:
\begin{equation} \begin{aligned}
{[s_1-1,s_2-1,s_3]}:\quad &\langle\mathbf{12}\rangle[\mathbf{12}]^{\Delta s_3-1}[\mathbf{23}]^{\Delta s_1}[\mathbf{31}]^{\Delta s_2},\\
[s_1,s_2-1,s_3-1]:\quad &\langle\mathbf{23}\rangle[\mathbf{12}]^{\Delta s_3}[\mathbf{23}]^{\Delta s_1-1}[\mathbf{31}]^{\Delta s_2},\\
[s_1-1,s_2,s_3-1]:\quad &\langle\mathbf{31}\rangle[\mathbf{12}]^{\Delta s_3}[\mathbf{23}]^{\Delta s_1}[\mathbf{31}]^{\Delta s_2-1}.\\
\end{aligned} \end{equation}
Performing this repeatedly, we find all $\prod_i 2\Delta s_i$ primary irreps.  The descendant irreps can be obtained by acting the $m_i$ or $\tilde{m}_i$ generators onto the primary irreps. From above we note the mass insertion gives $2s$-transversality irreps, and thus totally $\prod_i (2s_i+1)-1$ descendant irreps can be determined.


Let us consider $F\bar{F}S$ amplitudes. The primary irrep with maximal transversality is $[\mathbf{12}]$, and other primary irreps take $T^-_1 \cdot T^-_2 [\mathbf{12}]=\langle\mathbf{12}\rangle$ and $T^-_2 \cdot T^-_3 [\mathbf{12}]=T^-_3 \cdot T^-_1 [\mathbf{12}]=0$. All the descendant irreps are obtained as 
\begin{equation}
\includegraphics[width=0.9\linewidth,valign=c]{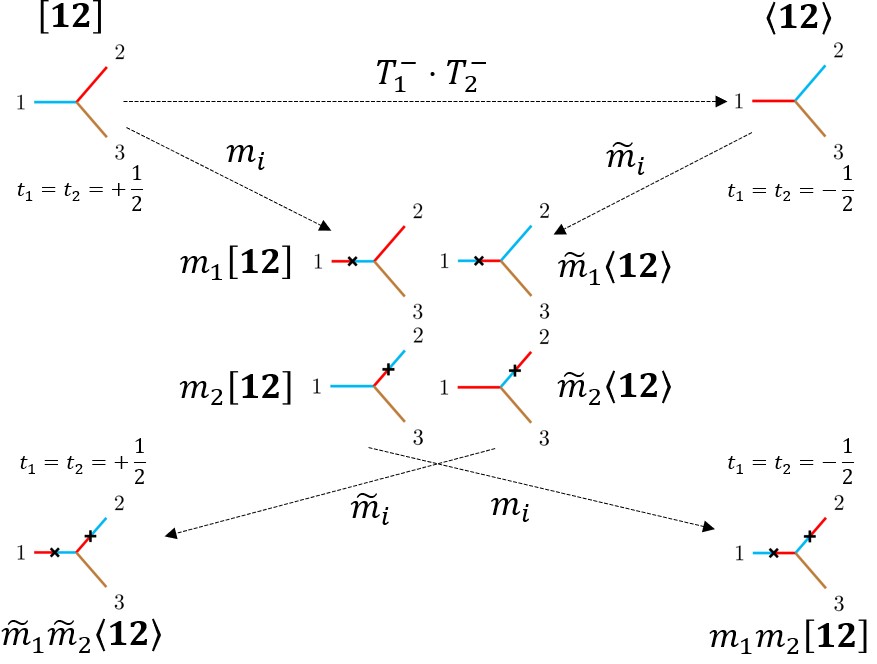} 
\end{equation}
Finally, we find the general form of the $F\bar{F}S$ amplitudes 
\begin{equation} \begin{aligned} \label{eq:resultFFS}
\mathcal{M}(F,\bar{F},S)
=&(c_1+c_2 \tilde{m}_1+c_3 \tilde{m}_2+c_4 \tilde{m}_1 \tilde{m}_2)\langle\mathbf{12}\rangle\\
&+(c_5+c_6 m_1+c_7 m_2+c_8 m_1 m_2)[\mathbf{12}],
\end{aligned} \end{equation}
where $c_i$ are independent coefficients for different $[t_1,t_2,t_3]$. This result keeps track of the chirality info while the AHH formalism does not have. The following example shows why transversality info is important. The charged pion decay $\pi^+\rightarrow \mu^+\nu_\mu$ and charged Higgs decay $H^+\rightarrow \mu^+\nu_\mu$ would give different results due to different transversality: $(t_1,t_2,t_3)=(0,+\frac{1}{2},-\frac{1}{2})$ verses $(0,-\frac{1}{2},-\frac{1}{2})$ respectively
\bea
\mathcal{M}_{\pi^+ \to \mu^+ \nu_\mu} = c_3 \tilde{m}_2\langle\mathbf{2}3\rangle, \ \ \mathcal{M}_{H^+\rightarrow \mu^+\nu_\mu} = c_1 \langle\mathbf{2}3\rangle. 
\eea
The appearance of mass insertion in the former but not the latter can be traced back to their different UV origins.

\begin{figure}[htbp]
\centering
\includegraphics[width=0.9\linewidth,valign=c]{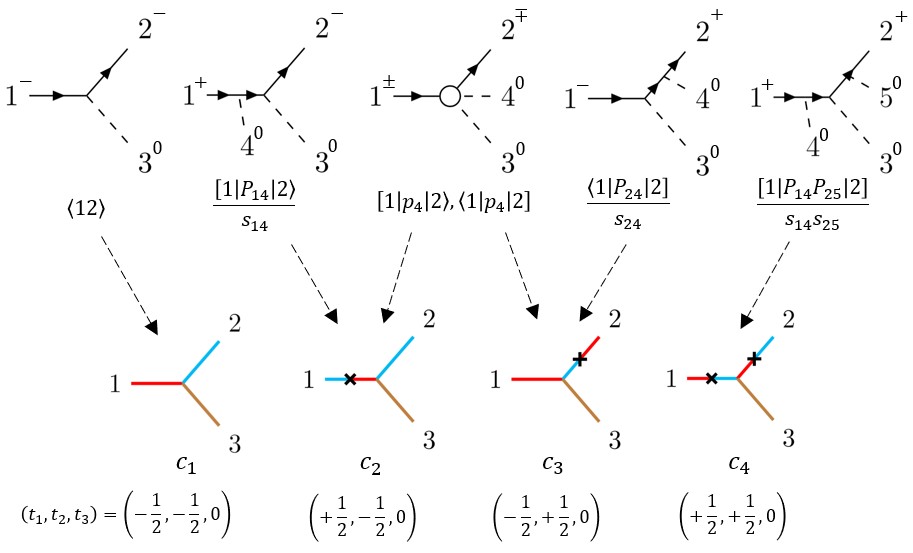}
\caption{The UV massless and IR massive amplitude correspondence for $F\bar{F}S$ amplitudes.  }
\label{fig:UV-FFS}
\end{figure}

In the electroweak theory with the Higgs symmetry breaking, massive amplitudes originate from UV massless helicity amplitudes. This connection has been discussed in the AHH formalism with the IR unification in Ref.~\cite{Christensen:2018zcq,Bachu:2019ehv,Liu:2022alx}. In this setup, the IR unification is not happen because IR massive amplitudes with different transversality are separable. In the high energy limit, the transversality and helicity for external particles in the amplitudes should be matched exactly: $t_i = h_i$, which we call "the chirality-helicity matching"~\footnote{This tells a correspondence between massive spin-transversality and massless helicity amplitudes. }. A massive momentum can be expanded using two massless spinors $\mathbf{p}_{\alpha\dot{\alpha}}  = \lambda_{\alpha} \tilde{\lambda}_{\dot{\alpha}} +\eta_{\alpha} \tilde{\eta}_{\dot{\alpha}}$, and thus the massive spinor can be expressed as the direct product
\begin{equation}
\begin{cases}
\lambda_{\alpha}^{I} = -\lambda_{\alpha} \zeta^{-I} +\eta_{\alpha} \zeta^{+I}, \\
\tilde{\lambda}_{\dot{\alpha}}^I = \tilde{\lambda}_{\dot{\alpha}} \zeta^{+I} +\tilde{\eta}_{\dot{\alpha}} \zeta^{-I},
\end{cases}
\label{eq:massless-decompose-su2-ahh}
\end{equation}
where $\zeta^{+}$ and $\zeta^{-}$ are two 2-dimensional vectors, $\lambda$ and $\eta$ are the massless spinors with transversality with $\lambda \sim \sqrt{2E}$ while $\eta \sim \frac{\mathbf{m}}{\sqrt{2E}}$ at the high energy limit $E\gg \mathbf{m}$.
Thus each term in Eq.~\eqref{eq:resultFFS} corresponds to different UV with helicity and transversality as shown in Figure~\ref{fig:UV-FFS}. 
From this, the charged Higgs decay should correspond to the 3-pt UV with scalar as Higgs boson, while the pion decay the 4-pt contact UV, in which the pion should be Goldstone. 

\begin{figure}[htbp]
\centering
\includegraphics[width=0.9\linewidth,valign=c]{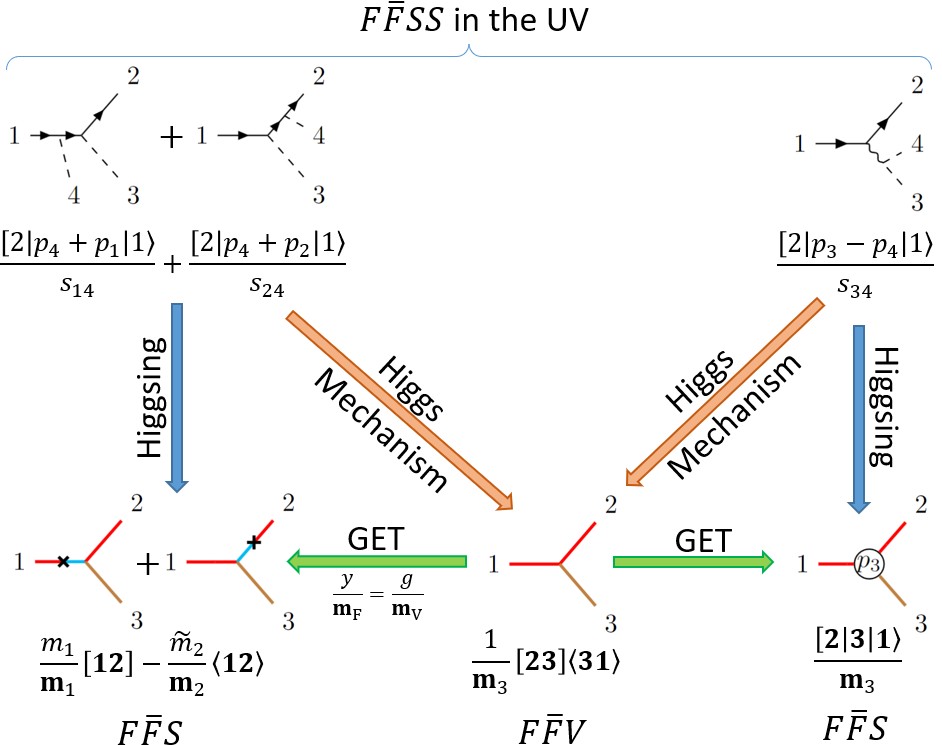}
\caption{The IR amplitudes with transversality $(t_1, t_2, t_3) = (-1/2,1/2,0)$ connect to the UV amplitudes $FFSS$ with helicity $(-,+,0)$. }
\label{fig:TopDown2}
\end{figure}

Let us switch to $F\bar{F}V$ amplitudes. 
As shown in Figure~\ref{fig:TopDown2}, it builds connection among concepts in the top-down and bottom-up understanding of the  electroweak theory:  
\begin{itemize}

\item \textit{On-shell Higgsing:} The IR mass insertion is from the Higgs boson VEV. Taking the on-shell limit~\cite{Balkin:2021dko} of the Higgs momentum $p_4 \to \eta_2$, the 4-pt massless $FFSS$ amplitude contributes to $F\bar{F}S$,
\begin{equation} 
\begin{tikzpicture}[scale=0.7,baseline=0.47cm] \begin{feynhand}
\vertex [particle] (i1) at (0,0.8) {$1$};
\vertex [particle] (i2) at (1.6,1.6) {$2$};
\vertex [particle] (i3) at (1.6,0) {$3$};
\vertex [particle] (i4) at (1.6,0.8) {$v$};
\vertex (v1) at (0.9,0.8);
\vertex (v2) at (0.9+0.7*0.33,0.8+0.8*0.33);
\graph{(i1)--[plain,red,very thick] (v1)};
\graph{(i2)--[plain,red,very thick](v2)--[plain,cyan,very thick] (v1)};
\graph{(i3)--[plain,brown,very thick] (v1)};
\draw[dash pattern=on 2 pt off 1 pt] (i4)--(v2);
\end{feynhand} \end{tikzpicture}
=\lim_{p_4\rightarrow\eta_2} v y^2\frac{\langle 1|P_{24}|2]}{s_{24}}
\rightarrow \frac{y}{\mathbf{m}_2}\tilde{m}_2\langle \mathbf{12}\rangle,
\end{equation}
where $y$ is the Yukawa coupling and the pole becomes fermion mass $s_{24}=(p_2+p_4)^2\rightarrow \mathbf{m}_2^2=y^2 v^2$. 

\item \textit{Higgs mechanism:} In the on-shell Higgsing, the transversality tells us the scalar in $F\bar{F}S$ is the Goldstone boson. If the Goldstone is eaten, then the massless $F\bar{F}SS$ is matched to the $F\bar{F}V$ amplitude  
\begin{equation} \label{eq:Higgsing}
\sum_{i=1}^2\lim_{p_4\rightarrow \eta_i}\frac{[2|P_{i4}|1\rangle}{s_{i4}}=\sum_{i=1}^2\frac{[2|\eta_i|1\rangle}{\mathbf{m}_i^2}\xrightarrow{\sum \eta=0} \frac{[\mathbf{23}]\langle \mathbf{31}\rangle}{\mathbf{m}_F^2}.
\end{equation}

\item \textit{Goldstone equivalence theorem (GET)}~\cite{Lee:1977eg,Chanowitz:1985hj}: Inversely, the $F\bar{F}V$ amplitude would relate to $F\bar{F}S$ amplitude by replacing 
\begin{equation} 
\mathcal{M}(F,\bar{F},V)\xRightarrow{\lambda^{I_1}_3 \tilde{\lambda}^{I_2}_3 \rightarrow p_3}\mathcal{M}(F,\bar{F},S),
\end{equation}
and identifying $y/\mathbf{m}_F = g/\mathbf{m}_V$.

\end{itemize}

\textbf{\textit{  Higher-point tree and loop amplitudes.}} From locality and unitarity, the factorization theorem tells how to construct higher point tree-level amplitudes by gluing the lower point $M^L$ and $M^R$ sub-amplitudes $\lim_{\mathbf{P}^2\rightarrow \mathbf{m}^2} (\mathbf{P}^2-\mathbf{m}^2) \mathcal{M}
=M^L\times M^R$, where $\mathbf{P}$ denotes momenta of on-shell propagator, with possible mass insertion in Eq.~\eqref{eq:onshellpropogator-momentum} and ~\eqref{eq:onshellpropogator-mass}.

\textit{1. tree-level 4-massive:} 
Consider the massive $b\bar{t}Wh$ amplitude. First choose the transversality $(-\frac{1}{2},+\frac{1}{2},0,0)$, and glue the 3-pt to give the $\mathbf{s}_{13}$-channel amplitude, which contains one mass insertion
\begin{equation} \label{eq:FFVS13}
\begin{tikzpicture}[baseline=0.8cm] \begin{feynhand}
\vertex [particle] (i1) at (0.375,1.55) {$1$};
\vertex [particle] (i2) at (0.375,0.25) {$2$};
\vertex [particle] (i3) at (1.425,0.25) {$4$};
\vertex [particle] (i4) at (1.425,1.55) {$3$};
\vertex (v1) at (0.9,0.6);
\vertex (v2) at (0.9,1.2);
\vertex (v3) at (0.9,0.9);
\graph{(i1)--[plain,red,very thick] (v2)--[plain,brown,very thick] (i4)};
\graph{(i2)--[plain,red,very thick] (v1)--[plain,brown,very thick] (i3)};
\graph{(v1)--[plain,cyan,very thick] (v3)--[plain,red,very thick] (v2)};
\draw[very thick] plot[mark=x,mark size=2.7] coordinates {(0.9,0.9)};
\end{feynhand} \end{tikzpicture}
=\langle\mathbf{13}\rangle[\mathbf{3}|_{\dot{\alpha}}\times\frac{\tilde{m}_t\delta_{\dot{\beta}}^{\dot{\alpha}} }{\mathbf{s}_{13}-\mathbf{m}_t^2}\times |\mathbf{2}]^{\dot{\beta}}
=\frac{\tilde{m}_t\langle\mathbf{13}\rangle[\mathbf{32}]}{\mathbf{s}_{13}-\mathbf{m}_t^2}.
\end{equation}
Summing over all the channels with all transversality, we obtain the final result ($m_b \to 0$)
\begin{equation} \begin{aligned}
\mathcal{M}
=&\mathcal{M}_{(-\frac{1}{2},+\frac{1}{2},0,0)}+\mathcal{M}_{(+\frac{1}{2},+\frac{1}{2},0,0)}+\mathcal{M}_{(-\frac{1}{2},-\frac{1}{2},0,0)}\\
=&
g^2 m_W\frac{2\mathbf{m}^2_W\langle\mathbf{13}\rangle[\mathbf{23}]+\langle\mathbf{1}|\mathbf{P}_{12}|\mathbf{2}]\langle\mathbf{3}|\mathbf{p}_4|\mathbf{3}]}{\mathbf{s}_{12}-\mathbf{m}_W^2}\\
&+y_t g\left(\frac{\tilde{m}_t\langle\mathbf{13}\rangle[\mathbf{32}]}{\mathbf{s}_{13}-\mathbf{m}_t^2}+\frac{\langle\mathbf{2}|\mathbf{P}_{13}|\mathbf{3}]\langle\mathbf{31}\rangle}{\mathbf{s}_{13}-\mathbf{m}_t^2}\right).\\
\end{aligned} \end{equation}

\textit{2. tree-level QED:}
Then we consider the 4-point amplitudes in the massive QED, in which the AHH formalism hardly obtains correct result~\cite{Christensen:2022nja,Lai:2023upa,Ema:2024vww}. For the scattering process $e\bar{e}\rightarrow\mu\bar{\mu}$, there is only one channel $s_{12}$ with photon propagator. Since the photon only couples to the spin-$\frac{1}{2}$ particle with same chirality, the transversality take $(\pm\frac{1}{2},\mp\frac{1}{2},\pm\frac{1}{2},\mp\frac{1}{2})$ and $(\pm\frac{1}{2},\mp\frac{1}{2},\mp\frac{1}{2},\pm\frac{1}{2})$, which can be given by the gluing technique. Diagrammatically, $(\pm\frac{1}{2},\mp\frac{1}{2},\pm\frac{1}{2},\mp\frac{1}{2})$ give
\begin{equation} \begin{aligned}
\begin{tikzpicture}[baseline=0.8cm] \begin{feynhand}
\setlength{\feynhandblobsize}{2mm}
\vertex [particle] (i1) at (0.25,1.525) {$1^{+}$};
\vertex [particle] (i2) at (0.2,0.3) {$2^{-}$};
\vertex [particle] (i3) at (2.00,0.3) {$3^{+}$};
\vertex [particle] (i4) at (1.95,1.525) {$4^{-}$};
\vertex (a1) at (0.6-0.6*0.33,0.9+0.9*0.33);
\vertex (a2) at (1.6+0.6*0.33,0.9-0.9*0.33);
\vertex (v1) at (0.6,0.9);
\vertex [ringblob,color=cyan,fill=white] (v2) at (0.9,0.9) {};
\vertex (v5) at (1.1,0.9);
\vertex [ringblob,color=red,fill=white] (v3) at (1.3,0.9) {};
\vertex (v4) at (1.6,0.9);
\graph{(i1)--[plain,cyan,very thick](v1)--[plain,cyan,very thick] (i2)};
\graph{(i4)--[plain,cyan,very thick](v4)--[plain,cyan,very thick](i3)};
\graph{(v1)--[plain,cyan](v2)--[plain,cyan](v5)--[plain,red,slash={[style=black]0}](v3)--[plain,red](v4)};
\end{feynhand} \end{tikzpicture}+
\begin{tikzpicture}[baseline=0.8cm] \begin{feynhand}
\setlength{\feynhanddotsize}{1mm}
\vertex [particle] (i1) at (0.25,1.525) {$1^{+}$};
\vertex [particle] (i2) at (0.2,0.3) {$2^{-}$};
\vertex [particle] (i3) at (2.00,0.3) {$3^{+}$};
\vertex [particle] (i4) at (1.95,1.525) {$4^{-}$};
\vertex (a1) at (0.6-0.6*0.33,0.9+0.9*0.33);
\vertex (a2) at (1.6+0.6*0.33,0.9-0.9*0.33);
\vertex (v1) at (0.6,0.9);
\vertex [crossdot,color=red,fill=white] (v2) at (0.9,0.9) {};
\vertex (v5) at (1.1,0.9);
\vertex [crossdot,color=cyan,fill=white] (v3) at (1.3,0.9) {};
\vertex (v4) at (1.6,0.9);
\graph{(i1)--[plain,red,very thick](v1)--[plain,red,very thick] (i2)};
\graph{(i4)--[plain,red,very thick](v4)--[plain,red,very thick](i3)};
\graph{(v1)--[plain,red](v2)--[plain,red](v5)--[plain,cyan,slash={[style=black]0}](v3)--[plain,cyan](v4)};
\end{feynhand} \end{tikzpicture}.
\end{aligned} \end{equation} 
Given that the minimal $F\bar{F}\gamma$ 3-pt amplitudes should be selected by the chirality-helicity matching which eliminates the unwanted UVs, the final results take the correct form
\begin{equation} \begin{aligned}
\mathcal{M}{e\bar{e}\to\mu\bar{\mu}}=\frac{[\mathbf{13}]\langle\mathbf{24}\rangle+[\mathbf{24}]\langle\mathbf{13}\rangle+[\mathbf{14}]\langle\mathbf{23}\rangle+[\mathbf{23}]\langle\mathbf{14}\rangle}{\mathbf{s}_{12}}.
\end{aligned} \end{equation}

\textit{3. loop-level 4-massive:} The integrand of the loop-level amplitude can be also calculated by the on-shell technique. Consider the 1-loop contribution in the Higgs decay channel $h\rightarrow b\bar{b}$. For simplicity, we focus on the one-particle irreducible (1PI) loop-level diagram with internal $W$ boson. 

In this case, the transversality are $(t_1,t_2,t_3)=(0,-\frac{1}{2},+\frac{1}{2})$. We can calculate the leading contribution for this diagram: it must have at least one mass insertion as
\begin{equation} \label{eq:hbb}
\begin{tikzpicture}[baseline=0.7cm] \begin{feynhand}
\setlength{\feynhandarrowsize}{3.5pt}
\vertex [particle] (i1) at (-0.15,0.8) {$h$};
\vertex [particle] (i2) at (1.7,1.6) {$b$};
\vertex [particle] (i3) at (1.7,0) {$\bar{b}$};
\vertex (v1) at (0.5,0.8);
\vertex (v2) at (1.1,1.2);
\vertex (v3) at (1.1,0.4);
\graph{(i1)--[sca] (v1)};
\graph{(v2)--[bos,edge label=$W$] (v3)};
\graph{(i3)--[fer] (v3)--[fer,edge label=$t$] (v1)--[fer,edge label=$t$] (v2)--[fer] (i2)};
\end{feynhand} \end{tikzpicture}
\Rightarrow
\begin{tikzpicture}[baseline=0.7cm] \begin{feynhand}
\vertex [particle] (i1) at (-0.15,0.8) {$1$};
\vertex [particle] (i2) at (1.7,1.6) {$2$};
\vertex [particle] (i3) at (1.7,0,0) {$3$};
\vertex (v1) at (0.5,0.8);
\vertex (v2) at (1.1,1.2);
\vertex (v3) at (1.1,0.4);
\vertex (v4) at (0.8,0.6);
\vertex (v5) at (0.8,1.0);
\graph{(i1)--[plain,brown,very thick] (v1)};
\graph{(v2)--[plain,red,very thick] (i2)};
\graph{(i3)--[plain,red,very thick] (v3)};
\graph{(v3)--[plain,red,very thick] (v4)--[plain,red,very thick] (v1)--[plain,cyan,very thick] (v5)--[plain,red,very thick] (v2)--[plain,brown,very thick] (v3)};
\draw[very thick] plot[mark=x,mark size=2.7,mark options={rotate=30}] coordinates {(v5)};
\path 
(0.65,1.3) node(A1) [rectangle,scale=0.8] {$l_2$}
(0.65,0.4) node(A2) [rectangle,scale=0.8] {$l_1$}
(1.3,0.8) node(A3) [rectangle,scale=0.8] {$l_3$};
\end{feynhand} \end{tikzpicture}+
\begin{tikzpicture}[baseline=0.7cm] \begin{feynhand}
\vertex [particle] (i1) at (-0.15,0.8) {$1$};
\vertex [particle] (i2) at (1.7,1.6) {$2$};
\vertex [particle] (i3) at (1.7,0) {$3$};
\vertex (v1) at (0.5,0.8);
\vertex (v2) at (1.1,1.2);
\vertex (v3) at (1.1,0.4);
\vertex (v4) at (0.8,0.6);
\vertex (v5) at (0.8,1.0);
\graph{(i1)--[plain,brown,very thick] (v1)};
\graph{(v2)--[plain,red,very thick] (i2)};
\graph{(i3)--[plain,red,very thick] (v3)};
\graph{(v3)--[plain,red,very thick] (v4)--[plain,cyan,very thick] (v1)--[plain,red,very thick] (v5)--[plain,red,very thick] (v2)--[plain,brown,very thick] (v3)};
\draw[very thick] plot[mark=x,mark size=2.7,mark options={rotate=30}] coordinates {(v4)};
\path 
(0.65,1.2) node(A1) [rectangle,scale=0.8] {$l_2$}
(0.65,0.3) node(A2) [rectangle,scale=0.8] {$l_1$}
(1.3,0.8) node(A3) [rectangle,scale=0.8] {$l_3$};
\end{feynhand} \end{tikzpicture},
\end{equation}
where the momenta $l_1$ and $l_2$ correspond to the top quark propagator, and the $l_3$ the $W$ propagator.

Using the unitarity cut~\cite{Bern:1994zx,Bern:1994cg,Britto:2004nc}, two propagators in the loop go on-shell, say $l_1$ and $l_3$. The 1-loop amplitude becomes an integral of two tree-level amplitudes as  
\begin{equation}
\text{Cut}\mathcal{M}^{\text{1PI}}=\int d\Omega
\mathcal{M}(l_3,2,1,l_1)\times \mathcal{M}(-l_1,3,-l_3),
\end{equation}
where $d\Omega$ is the solid angle. Combining with Eq.~\eqref{eq:hbb}, the leading contributions of the cut diagram are
\bea \left(\frac{\tilde{m}_t\langle\mathbf{2}l_3^{(I_1}\rangle[l_3^{I_2)} l_1^J]}{l_2^2-\mathbf{m}_t^2}
+\frac{\langle\mathbf{2} l_3^{(I_1}\rangle[l_3^{I_2)}|l_2|l_1^{J}\rangle }{l_2^2-\mathbf{m}_t^2}\right)\langle l_{1J} l_{3 I_1}\rangle[l_{3 I_2} \mathbf{3}],
 \eea
where $l_2=\mathbf{p}_2+l_3=\mathbf{p}_1+l_1$. After some calculation, we obtain the correct integrand.

\textbf{\textit{  Summary and Outlook.}} The on-shell method of scattering amplitudes has the advantage of all the space-time symmetries manifest. We recognize the maximal spacetime symmetry for massive amplitudes with complex masses. This maximal symmetry determines both the general spin-transversality spinors for spin-$s$ particle and the 3-pt massive amplitudes with chirality flip. Thus while the AHH formalism could describe the Lorentz structure of massive amplitudes, this symmetry further determines the chirality structure, which describes processes with mass enhancements. Furthermore, in the case different UVs need to be disentangled, the chirality-helicity unification in the high energy limit could help to do so. We find this disentanglement is crucial to obtain the correct results of the $e^+e^- \to \mu^+\mu^-$ process.


There are many research directions to be explored using the spin-transversality spinors. {\it i)}. There is deep connection between the massive spinors and 6d massless spinors~\cite{Cheung:2009dc,Boels:2009bv,Bern:2010qa,Davies:2011vt,AccettulliHuber:2019abj}. Although these new spinors do not have $SO(5,1)$ symmetry, it still can be implemented as four dimensional reduction of six dimensional massless amplitudes. {\it ii)}. Compared to AHH formalism, the transversality-helicity connection at the UV and IR builds a correspondence between massless helicity and massive amplitudes. Applying this to electroweak amplitudes would give a better understanding on the Higgs mechanism, Goldstone equivalence theorem, and perturbative unitarity, etc. {\it iii)}. The transversality and chirality identification provides a phenomenological connection to the massive helicity amplitudes used in electroweak calculation~\cite{Hagiwara:1985yu,Murayama:1992gi}. {\it iv)}. The on-shell technique developed here for chiral-eigenstates amplitudes provides a new way of calculating the loop diagrams with mass insertion and chirality flip. {\it v)}. the above discussion could extend to higher spin particles, which provides more insight on double copy relations~\cite{Kawai:1985xq,Bern:2008qj,Bern:2019prr}, gravitational wave calculations~\cite{Guevara:2018wpp,Chung:2018kqs,Bern:2019nnu,Maybee:2019jus}, etc.  All of these deserve further investigation in future.

\begin{acknowledgments}


\textbf{\textit{Acknowledgements.}} 
We thank Bo Feng, Song He, Xiao-Gang He, Mingxing Luo and Gang Yang for their valuable discussions. This work is supported by the National Science Foundation of China under Grants No. 12347105, No. 12375099 and No. 12047503, and the National Key Research and Development Program of China Grant No. 2020YFC2201501, No. 2021YFA0718304.

\end{acknowledgments}

\bibliographystyle{h-physrev} 
\bibliography{refs.bib}

\appendix

\begin{widetext}

\par\noindent\rule{\textwidth}{0.8pt}


\section{Supplemental Material for ``Extended Poincare Symmetry dictates Massive Scattering Amplitudes"}

\section{1. Massive $F\bar{F}V$ and $F\bar{F}VS$ in the Electroweak Theory}

The massive $F\bar{F}V$ amplitudes can also be determined by extended Poincare symmetry. There are four primary reps:
\begin{equation} \begin{aligned} \label{eq:FFV}
\begin{tikzpicture}[baseline=0.7cm] \begin{feynhand}
\vertex [particle] (i1) at (0,0.8) {$1$};
\vertex [particle] (i2) at (1.6,1.6) {$2$};
\vertex [particle] (i3) at (1.6,0) {$3$};
\vertex (v1) at (0.9,0.8);
\graph{(i1)--[plain,red,very thick] (v1)};
\graph{(i2)--[plain,red,very thick] (v1)};
\graph{(i3)--[plain,brown,very thick] (v1)};
\end{feynhand} \end{tikzpicture}={[\mathbf{23}]\langle\mathbf{31}\rangle},\quad
\begin{tikzpicture}[baseline=0.7cm] \begin{feynhand}
\vertex [particle] (i1) at (0,0.8) {$1$};
\vertex [particle] (i2) at (1.6,1.6) {$2$};
\vertex [particle] (i3) at (1.6,0) {$3$};
\vertex (v1) at (0.9,0.8);
\graph{(i1)--[plain,cyan,very thick] (v1)};
\graph{(i2)--[plain,cyan,very thick] (v1)};
\graph{(i3)--[plain,brown,very thick] (v1)};
\end{feynhand} \end{tikzpicture}={\langle\mathbf{23}\rangle[\mathbf{31}]},\quad
\begin{tikzpicture}[baseline=0.7cm] \begin{feynhand}
\vertex [particle] (i1) at (0,0.8) {$1$};
\vertex [particle] (i2) at (1.6,1.6) {$2$};
\vertex [particle] (i3) at (1.6,0) {$3$};
\vertex (v1) at (0.9,0.8);
\graph{(i1)--[plain,cyan,very thick] (v1)};
\graph{(i2)--[plain,red,very thick] (v1)};
\graph{(i3)--[plain,cyan,very thick] (v1)};
\end{feynhand} \end{tikzpicture}=[\mathbf{23}][\mathbf{31}],\quad
\begin{tikzpicture}[baseline=0.7cm] \begin{feynhand}
\vertex [particle] (i1) at (0,0.8) {$1$};
\vertex [particle] (i2) at (1.6,1.6) {$2$};
\vertex [particle] (i3) at (1.6,0) {$3$};
\vertex (v1) at (0.9,0.8);
\graph{(i1)--[plain,red,very thick] (v1)};
\graph{(i2)--[plain,cyan,very thick] (v1)};
\graph{(i3)--[plain,red,very thick] (v1)};
\end{feynhand} \end{tikzpicture}=
\langle\mathbf{23}\rangle\langle\mathbf{31}\rangle.
\end{aligned} \end{equation}
The first two reps correspond to the renormalisable minimal couplings, while the last two correspond to the effective operator $X F\bar{F} D$.

Then performing the mass insertion on the primary reps, we obtain the descendant reps. For $[\mathbf{23}]\langle\mathbf{31}\rangle$, one mass insertion gives
\begin{equation} \begin{aligned}
\begin{tikzpicture}[baseline=0.7cm] \begin{feynhand}
\vertex [particle] (i1) at (0,0.8) {$1$};
\vertex [particle] (i2) at (1.6,1.6) {$2$};
\vertex [particle] (i3) at (1.6,0) {$3$};
\vertex (v1) at (0.9,0.8);
\vertex (v2) at (0.6,0.8);
\graph{(i1)--[plain,cyan,very thick](v2)--[plain,red,very thick] (v1)};
\graph{(i2)--[plain,red,very thick] (v1)};
\graph{(i3)--[plain,brown,very thick] (v1)};
\draw[very thick] plot[mark=x,mark size=2.7,mark options={rotate=0}] coordinates {(v2)};
\end{feynhand} \end{tikzpicture}=\tilde{m}_1[\mathbf{23}]\langle\mathbf{31}\rangle,
\begin{tikzpicture}[baseline=0.7cm] \begin{feynhand}
\vertex [particle] (i1) at (0,0.8) {$1$};
\vertex [particle] (i2) at (1.6,1.6) {$2$};
\vertex [particle] (i3) at (1.6,0) {$3$};
\vertex (v1) at (0.9,0.8);
\vertex (v2) at (0.9+0.7*0.35,0.8+0.8*0.35);
\graph{(i1)--[plain,red,very thick] (v1)};
\graph{(i2)--[plain,cyan,very thick](v2)--[plain,red,very thick] (v1)};
\graph{(i3)--[plain,brown,very thick] (v1)};
\draw[very thick] plot[mark=x,mark size=2.7,mark options={rotate=45}] coordinates {(v2)};
\end{feynhand} \end{tikzpicture}=m_2[\mathbf{23}]\langle\mathbf{31}\rangle,
\begin{tikzpicture}[baseline=0.7cm] \begin{feynhand}
\vertex [particle] (i1) at (0,0.8) {$1$};
\vertex [particle] (i2) at (1.6,1.6) {$2$};
\vertex [particle] (i3) at (1.6,0) {$3$};
\vertex (v1) at (0.9,0.8);
\vertex (v2) at (0.9+0.7*0.35,0.8-0.8*0.35);
\graph{(i1)--[plain,red,very thick] (v1)};
\graph{(i2)--[plain,red,very thick] (v1)};
\graph{(i3)--[plain,cyan,very thick](v2)--[plain,brown,very thick] (v1)};
\draw[very thick] plot[mark=x,mark size=2.7,mark options={rotate=45}] coordinates {(v2)};
\end{feynhand} \end{tikzpicture}=\tilde{m}_3[\mathbf{23}]\langle\mathbf{31}\rangle,
\begin{tikzpicture}[baseline=0.7cm] \begin{feynhand}
\vertex [particle] (i1) at (0,0.8) {$1$};
\vertex [particle] (i2) at (1.6,1.6) {$2$};
\vertex [particle] (i3) at (1.6,0) {$3$};
\vertex (v1) at (0.9,0.8);
\vertex (v2) at (0.9+0.7*0.35,0.8-0.8*0.35);
\graph{(i1)--[plain,red,very thick] (v1)};
\graph{(i2)--[plain,red,very thick] (v1)};
\graph{(i3)--[plain,red,very thick](v2)--[plain,brown,very thick] (v1)};
\draw[very thick] plot[mark=x,mark size=2.7,mark options={rotate=45}] coordinates {(v2)};
\end{feynhand} \end{tikzpicture}=m_3[\mathbf{23}]\langle\mathbf{31}\rangle,
\end{aligned} \end{equation}
Similarly, we can perform the mass insertion to other primary reps. After performing mass insertion repeatedly, we obtain the general form of the $F\bar{F}V$ amplitudes,
\begin{equation} \begin{aligned}
&\mathcal{M}(F,\bar{F},V)\\
=&(c_1+c_2 \tilde{m}_1+c_3 \tilde{m}_2+c_4 \tilde{m}_3+c_5 \tilde{m}_1 \tilde{m}_2+c_6 \tilde{m}_1 \tilde{m}_3+c_7 \tilde{m}_2 \tilde{m}_3+c_8 \tilde{m}_3^2+c_9 \tilde{m}_1 \tilde{m}_2 \tilde{m}_3+c_{10} \tilde{m}_1 \tilde{m}_3^2+c_{11} \tilde{m}_2 \tilde{m}_3^2+c_{12} \tilde{m}_1 \tilde{m}_2 \tilde{m}_3^2)\\
&\times\langle\mathbf{23}\rangle\langle\mathbf{31}\rangle
+(c_{13}+c_{14} \tilde{m}_1+c_{15} m_2+c_{16} \tilde{m}_3+c_{17} m_3+c_{18} \tilde{m}_1 m_2+c_{19} \tilde{m}_1 \tilde{m}_3+c_{20} \tilde{m}_1 m_3+c_{21} m_2 \tilde{m}_3+c_{22} m_2 m_3\\
&+c_{23} \tilde{m}_1 m_2 \tilde{m}_3+c_{24} \tilde{m}_1 m_2 m_3)[\mathbf{23}]\langle\mathbf{31}\rangle+(c_{25}+c_{26} m_1+c_{27} \tilde{m}_2+c_{28} \tilde{m}_3+c_{29} m_3+c_{30} m_1 \tilde{m}_2+c_{31} m_1 \tilde{m}_3+c_{32} m_1 m_3\\
&+c_{33} \tilde{m}_2 \tilde{m}_3+c_{34} \tilde{m}_2 m_3+c_{35} m_1 \tilde{m}_2 \tilde{m}_3+c_{36} m_1 \tilde{m}_2 m_3)\langle\mathbf{23}\rangle[\mathbf{31}]+(c_{37}+c_{38} m_1+c_{39} m_2+c_{40} \tilde{m}_3+c_{41} m_1 m_2+c_{42} m_1 \tilde{m}_3\\
&+c_{43} m_2 \tilde{m}_3+c_{44} \tilde{m}_3^2+c_{45} m_1 m_2 \tilde{m}_3+c_{46} m_1 \tilde{m}_3^2+c_{47} m_2 \tilde{m}_3^2+c_{48} m_1 m_2 \tilde{m}_3^2)[\mathbf{23}][\mathbf{31}].\\
\end{aligned} \end{equation}
Similar to $F\bar{F}S$, each coefficient in the above corresponds to a certain kind of UVs. In the top-down matching, the helicity of each UV particle should match the transversality of the corresponding IR particle, except for the extra Higgs boson: $t_i = h_i$ and thus $\sum h_i=\sum t_i$. 


\begin{figure}[htbp]
\centering
\includegraphics[width=0.8\linewidth,valign=c]{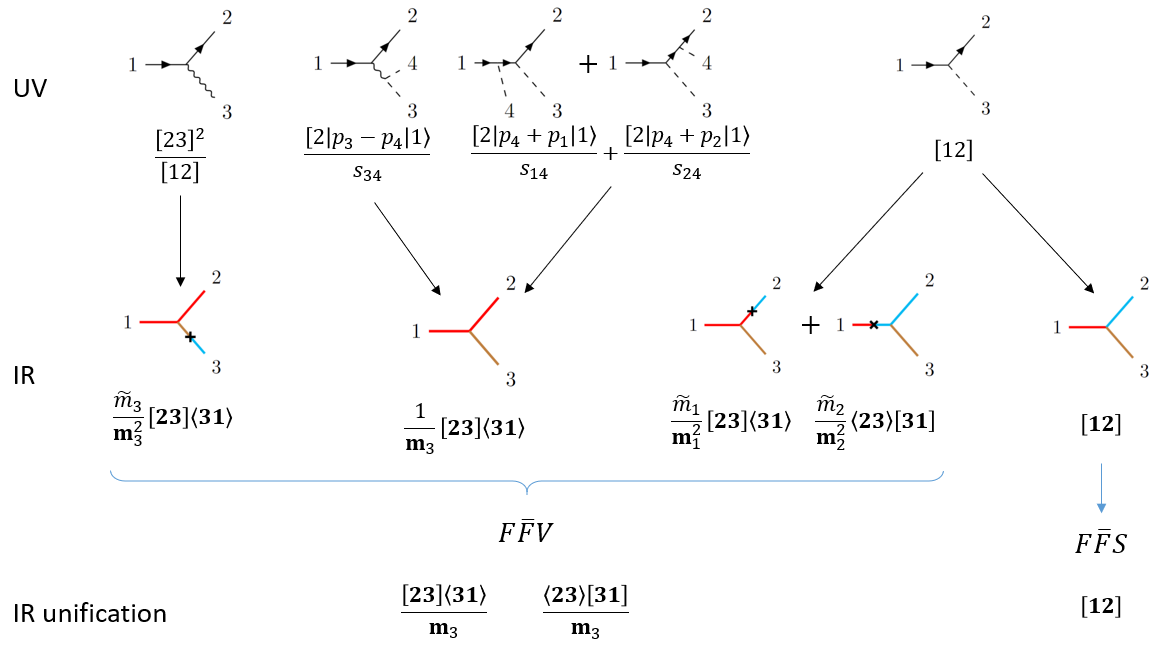}
\caption{The UV massless and IR massive amplitude correspondence for $F\bar{F}V$ amplitudes. The IR unification is also shown as connection to AHH formalism. }
\label{fig:TopDown}
\end{figure}

In the renormalizable theory, the 3-pt massless amplitude gives $\sum h_i=\pm 1$. Since the total transversality of the primary reps for $F\bar{F}V$ amplitude is $\sum t_i=0$ or $\pm2$, the 3-pt massless amplitudes must match to descendant reps, as shown in figure~\ref{fig:TopDown}. The massless $F\bar{F}V$ amplitudes match to the rep with mass insertion $\tilde{m}_3$, while massless $F\bar{F}S$ amplitudes match to the ones with $\tilde{m}_1$ and $\tilde{m}_2$, which gives the on-shell version of the Higgs mechanism. The coefficients $b_i$ with the constraint $b_1+b_2=1$ are determined by matching from massless UV. On the other hand, the 4-pt massless amplitude gives $\sum h_i=0$, so it matches to the primary rep.

In the IR, the massive $F\bar{F}V$ amplitudes would be unified into one form in the AHH formalism, when taking the masses to be real, and thus no transversality info. 
%
Note that the UV $F\bar{F}V$ and $F\bar{F}S$, give the result with different mass insertions. We can choose one as the standard form of the IR amplitude, e.g. the rep with mass insertion $\tilde{m}_3$. Other mass insertion $\tilde{m}_1$ and $\tilde{m}_2$ will tell us the relation between the coefficient $c_i$ and the Yukawa coupling in the UV.

Now we turn to calculate the 4-pt $b\bar{t}Wh$ amplitude. For simplicity, we only use the 3-pt amplitudes in the primary rep. Gluing the $F\bar{F}V$ amplitude in Eq.~\eqref{eq:FFV} and the $F\bar{F}S$ amplitude, we obtain the amplitudes in the $\mathbf{s}_{13}$ channel. This channel has two contributions with transversality $(-\frac{1}{2},\pm\frac{1}{2},0,0)$:
\bea
\begin{tikzpicture}[baseline=0.8cm] \begin{feynhand}
\vertex [particle] (i1) at (0.375,1.55) {$1$};
\vertex [particle] (i2) at (0.375,0.25) {$2$};
\vertex [particle] (i3) at (1.425,0.25) {$4$};
\vertex [particle] (i4) at (1.425,1.55) {$3$};
\vertex (v1) at (0.9,0.6);
\vertex (v2) at (0.9,1.2);
\vertex (v3) at (0.9,0.9);
\graph{(i1)--[plain,red,very thick] (v2)--[plain,brown,very thick] (i4)};
\graph{(i2)--[plain,red,very thick] (v1)--[plain,brown,very thick] (i3)};
\graph{(v1)--[plain,cyan,very thick] (v3)--[plain,red,very thick] (v2)};
\draw[very thick] plot[mark=x,mark size=2.7] coordinates {(0.9,0.9)};
\end{feynhand} \end{tikzpicture}
&=&\langle\mathbf{13}\rangle[\mathbf{3}|_{\dot{\alpha}}\times\frac{\tilde{m}_t\delta_{\dot{\beta}}^{\dot{\alpha}} }{\mathbf{s}_{13}-\mathbf{m}_t^2}\times |\mathbf{2}]^{\dot{\beta}}
=\frac{\tilde{m}_t\langle\mathbf{13}\rangle[\mathbf{32}]}{\mathbf{s}_{13}-\mathbf{m}_t^2},\\
\begin{tikzpicture}[baseline=0.8cm] \begin{feynhand}
\vertex [particle] (i1) at (0.375,1.55) {$1$};
\vertex [particle] (i2) at (0.375,0.25) {$2$};
\vertex [particle] (i3) at (1.425,0.25) {$4$};
\vertex [particle] (i4) at (1.425,1.55) {$3$};
\vertex (v1) at (0.9,0.6);
\vertex (v2) at (0.9,1.2);
\vertex (v3) at (0.9,0.9);
\graph{(i1)--[plain,red,very thick] (v2)--[plain,brown,very thick] (i4)};
\graph{(i2)--[plain,cyan,very thick] (v1)--[plain,brown,very thick] (i3)};
\graph{(v1)--[plain,red,very thick] (v3)--[plain,red,very thick] (v2)};
\end{feynhand} \end{tikzpicture}
&=&\langle\mathbf{13}\rangle[\mathbf{3}|_{\dot{\alpha}}\times\frac{\mathbf{P}_{\beta}^{\dot{\alpha}} }{\mathbf{s}_{13}-\mathbf{m}_t^2}\times |\mathbf{2}\rangle^{\beta}
=\frac{\langle\mathbf{2}|\mathbf{P}_{13}|\mathbf{3}]\langle\mathbf{31}\rangle}{\mathbf{s}_{13}-\mathbf{m}_t^2}.
\eea
The calculation for $\mathbf{s}_{14}$ channel is similar to the $\mathbf{s}_{13}$ channel, but the transversality are $(\mp\frac{1}{2},+\frac{1}{2},0,0)$. 

The $\mathbf{s}_{12}$ channel only has one contribution with $(-\frac{1}{2},+\frac{1}{2},0,0)$, which originates from gluing the $F\bar{F}V$ and $VVS$ amplitude. It contains the internal vector boson without mass insertion
\begin{equation}
\begin{tikzpicture}[baseline=0.8cm] \begin{feynhand}
\vertex [particle] (i1) at (0.25,1.425) {$1$};
\vertex [particle] (i2) at (0.25,0.375) {$2$};
\vertex [particle] (i3) at (1.55,0.375) {$3$};
\vertex [particle] (i4) at (1.55,1.425) {$4$};
\vertex (v1) at (0.6,0.9);
\vertex (v2) at (1.2,0.9);
\graph{(i1)--[plain,red,very thick] (v1)--[plain,red,very thick] (i2)};
\graph{(i4)--[plain,brown,very thick] (v2)--[plain,brown,very thick] (i3)};
\graph{(v1)--[plain,brown,very thick] (v2)};
\end{feynhand} \end{tikzpicture}.
\end{equation}
We can still use the extended Poincar\'e symmetry to study the internal vector boson by gluing the external vector and antivector boson with different transversality 
\begin{equation} \begin{aligned}
\begin{pmatrix}
\begin{tikzpicture}[baseline=0.7cm] \begin{feynhand}
\vertex [dot] (i1) at (1.6,0.8) {};
\vertex [dot] (v1) at (0,0.8) {};
\vertex (v2) at (0.53,0.8);
\vertex (v3) at (1.07,0.8);
\graph{(v1)--[plain,red,very thick] (v2)--[plain,brown,very thick] (v3)--[plain,cyan,very thick] (i1)};
\draw[very thick] plot[mark=x,mark size=2.7] coordinates {(v2)};
\draw[very thick] plot[mark=x,mark size=2.7] coordinates {(v3)};
\end{feynhand} \end{tikzpicture} &
\begin{tikzpicture}[baseline=0.7cm] \begin{feynhand}
\vertex [dot] (v1) at (0,0.8) {};
\vertex [dot] (v2) at (1.6,0.8) {};
\vertex (v3) at (0.8,0.8);
\graph{(v2)--[plain,cyan,very thick] (v3)--[plain,brown,very thick] (v1)};
\draw[very thick] plot[mark=x,mark size=2.7] coordinates {(v3)};
\end{feynhand} \end{tikzpicture} &
\begin{tikzpicture}[baseline=0.7cm] \begin{feynhand}
\vertex [dot] (v1) at (0,0.8) {};
\vertex [dot] (v2) at (1.6,0.8) {};
\vertex (v3) at (0.8,0.8);
\graph{(v2)--[plain,cyan,very thick] (v3)--[plain,cyan,very thick] (v1)};
\end{feynhand} \end{tikzpicture} \\
\begin{tikzpicture}[baseline=0.7cm] \begin{feynhand}
\vertex [dot] (v1) at (0,0.8) {};
\vertex [dot] (v2) at (1.6,0.8) {};
\vertex (v3) at (0.8,0.8);
\graph{(v2)--[plain,brown,very thick] (v3)--[plain,red,very thick] (v1)};
\draw[very thick] plot[mark=x,mark size=2.7] coordinates {(v3)};
\end{feynhand} \end{tikzpicture} & 
\begin{tikzpicture}[baseline=0.7cm] \begin{feynhand}
\vertex [dot] (v1) at (0,0.8) {};
\vertex [dot] (v2) at (1.6,0.8) {};
\graph{(v2)--[plain,brown,very thick] (v1)};
\end{feynhand} \end{tikzpicture} & 
\begin{tikzpicture}[baseline=0.7cm] \begin{feynhand}
\vertex [dot] (v1) at (0,0.8) {};
\vertex [dot] (v2) at (1.6,0.8) {};
\vertex (v3) at (0.8,0.8);
\graph{(v2)--[plain,brown,very thick] (v3)--[plain,cyan,very thick] (v1)};
\draw[very thick] plot[mark=x,mark size=2.7] coordinates {(v3)};
\end{feynhand} \end{tikzpicture} \\
\begin{tikzpicture}[baseline=0.7cm] \begin{feynhand}
\vertex [dot] (v1) at (0,0.8) {};
\vertex [dot] (v2) at (1.6,0.8) {};
\vertex (v3) at (0.8,0.8);s
\graph{(v2)--[plain,red,very thick] (v3)--[plain,red,very thick] (v1)};
\end{feynhand} \end{tikzpicture} & 
\begin{tikzpicture}[baseline=0.7cm] \begin{feynhand}
\vertex [dot] (v1) at (0,0.8) {};
\vertex [dot] (v2) at (1.6,0.8) {};
\vertex (v3) at (0.8,0.8);
\graph{(v2)--[plain,red,very thick] (v3)--[plain,brown,very thick] (v1)};
\draw[very thick] plot[mark=x,mark size=2.7] coordinates {(v3)};
\end{feynhand} \end{tikzpicture} & 
\begin{tikzpicture}[baseline=0.7cm] \begin{feynhand}
\vertex [dot] (i1) at (1.6,0.8) {};
\vertex [dot] (v1) at (0,0.8) {};
\vertex (v2) at (0.53,0.8);
\vertex (v3) at (1.07,0.8);
\graph{(v1)--[plain,cyan,very thick] (v2)--[plain,brown,very thick] (v3)--[plain,red,very thick] (i1)};
\draw[very thick] plot[mark=x,mark size=2.7] coordinates {(v2)};
\draw[very thick] plot[mark=x,mark size=2.7] coordinates {(v3)};
\end{feynhand} \end{tikzpicture} \\
\end{pmatrix}=
\begin{pmatrix}
m^2\delta_{(\alpha_1}^{(\beta_1}\delta_{\alpha_2)}^{\beta_2)} & m\mathbf{p}_{\dot{\alpha}_2}^{(\beta_2}\delta_{\alpha_1}^{\beta_1)} & \mathbf{p}_{(\dot{\alpha}_1}^{(\beta_1}\mathbf{p}_{\dot{\alpha}_2)}^{\beta_2)} \\
m\mathbf{p}_{(\alpha_1}^{\dot{\beta}_1}\delta_{\alpha_2)}^{\beta_2} & 2\mathbf{m}^2\delta_{\alpha_1}^{\beta_1}\delta_{\dot{\alpha}_2}^{\dot{\beta}_2}-\mathbf{p}_{\dot{\alpha}_2}^{\beta_1}\mathbf{p}_{\alpha_1}^{\dot{\beta}_2} & \tilde{m}\mathbf{p}_{(\dot{\alpha}_1}^{\beta_1}\delta_{\dot{\alpha}_2)}^{\dot{\beta}_2} \\
\mathbf{p}_{(\alpha_1}^{(\dot{\beta}_1}\mathbf{p}_{\alpha_2)}^{\dot{\beta}_2)} & \tilde{m}\mathbf{p}_{\alpha_1}^{(\dot{\beta}_1}\delta_{\dot{\alpha}_2}^{\dot{\beta}_2)} & \tilde{m}^2\delta_{(\dot{\alpha}_1}^{(\dot{\beta}_2}\delta_{\dot{\alpha}_2)}^{\dot{\beta}_2)} \\
\end{pmatrix},
\end{aligned} \end{equation}
where $(\cdots)$ stands for symmetrization of indices. Note that the mass insertion is given by crosses explicitly. Now we can calculate the $\mathbf{s}_{12}$ channel
\begin{equation}
\langle\mathbf{1}|_{\beta_1}[\mathbf{2}|^{\dot{\alpha}_2}\times\frac{2\mathbf{m}^2_W\delta_{\alpha_1}^{\beta_1}\delta_{\dot{\alpha}_2}^{\dot{\beta}_2}-\mathbf{P}_{\dot{\alpha}_2}^{\beta_1}\mathbf{P}_{\alpha_1}^{\dot{\beta}_2}}{\mathbf{s}_{12}-\mathbf{m}_W^2}\times |\mathbf{3}]_{\dot{\beta}_2}|\mathbf{3}\rangle^{\alpha_1}
=\frac{2\mathbf{m}^2_W\langle\mathbf{13}\rangle[\mathbf{23}]+\langle\mathbf{1}|\mathbf{P}_{12}|\mathbf{2}]\langle\mathbf{3}|\mathbf{p}_4|\mathbf{3}]}{\mathbf{s}_{12}-\mathbf{m}_W^2},
\end{equation}
where $|\mathbf{3}]_{\dot{\beta}_2}|\mathbf{3}\rangle^{\alpha_1}$ is wavefunction of the vector boson. Summing over amplitudes with all transversality, we obtain the final result
\begin{equation} \begin{aligned}
\mathcal{M}
=&\mathcal{M}_{(-\frac{1}{2},+\frac{1}{2},0,0)}+\mathcal{M}_{(+\frac{1}{2},+\frac{1}{2},0,0)}+\mathcal{M}_{(-\frac{1}{2},-\frac{1}{2},0,0)}\\
=&g^2 m_W\frac{2\mathbf{m}^2_W\langle\mathbf{13}\rangle[\mathbf{23}]+\langle\mathbf{1}|\mathbf{P}_{12}|\mathbf{2}]\langle\mathbf{3}|\mathbf{p}_4|\mathbf{3}]}{\mathbf{s}_{12}-\mathbf{m}_W^2}+y_t g\left(\frac{\tilde{m}_t\langle\mathbf{13}\rangle[\mathbf{32}]}{\mathbf{s}_{13}-\mathbf{m}_t^2}+\frac{\langle\mathbf{2}|\mathbf{P}_{13}|\mathbf{3}]\langle\mathbf{31}\rangle}{\mathbf{s}_{13}-\mathbf{m}_t^2}\right)+y_b g\left(\frac{\tilde{m}_b\langle\mathbf{13}\rangle[\mathbf{32}]}{\mathbf{s}_{14}-\mathbf{m}_b^2}+\frac{[\mathbf{2}|\mathbf{P}_{14}|\mathbf{3}\rangle[\mathbf{31}]}{\mathbf{s}_{14}-\mathbf{m}_b^2}\right).\\
\end{aligned} \end{equation}

\section{2. 1-massless 2-massive amplitude} 
Consider another type of 3-pt amplitudes: 1-massless 2-massive amplitudes. For massless particle, the chirality is equal to the helicity. There is no mass insertion for massless spinors, so the transversality is also the same as the helicity. Therefore, we can still use transversality $(t_1,t_2,t_3)$ to label the reps of 1-massless 2-massive amplitudes. 

Suppose that particle 3 is a massless particle with positive helicity $h_3\ge s_2 \ge s_1$ and the two massive particle have different masses. In this case, the wave function of particle 3 is $\lambda_3^{h_3}$. If  $s_1+s_2\ge h_3$ is satisfied \footnote{If $s_1+s_2<h_3$, the amplitude does not exist. However, the amplitudes with $s_2+h_3\ge s_1$ or $s_1+h_3\ge s_2$ are still constructible by considering the non-trivial internal structure, i.e. the monomial in $p$.}, we can choose the primary rep to be 
\begin{equation} \label{eq:HighAmp}
[\mathbf{12}]^{s_1+s_2-h_3}[\mathbf{2}3]^{s_2-s_1+h_3}[3\mathbf{1}]^{s_1-s_2+h_3},
\end{equation}
and then we use the same procedure as the 3-massive amplitude to derive other reps.

When two masses are equal, we have an additional constraint
$\langle3|\mathbf{p}_1|3]=\mathbf{m}_2^2-\mathbf{m}_1^2=0$. Therefore, $\lambda_3^I$ should be proportional to $\mathbf{p}_1 \tilde{\lambda}_3^I$. Their proportionality gives the $\mathbf{x}$-factor. There are some possibilities for the $\mathbf{x}$-factor, each carrying different transversality, defined as
\begin{equation} \label{eq:xfactor2}
\mathbf{x}\equiv\left\{\frac{[3|\mathbf{p}_1}{\mathbf{m}\langle3|},\frac{[3|\mathbf{p}_1}{m_1\langle3|},\frac{[3|\mathbf{p}_1}{m_2\langle3|},\frac{[3|\mathbf{p}_1}{\tilde{m}_1\langle3|},\frac{[3|\mathbf{p}_1}{\tilde{m}_2\langle3|}\right\}
\end{equation}
where $\mathbf{m}=\mathbf{m}_1=\mathbf{m}_2$. For our purposes, we prefer the first one with $(t_1,t_2,t_3)=(0,0,+1)$, because it will not affect the chirality flip of particles 1 and 2. So we take the $\mathbf{x}$-factor to be
\begin{equation} \label{eq:xfactor}
\mathbf{x}=\frac{[3|\mathbf{p}_1|\xi\rangle}{\mathbf{m}\langle3\xi\rangle} = \epsilon \cdot \frac{\mathbf{p}_1}{\mathbf{m}},
\end{equation}
where $|\xi\rangle$ is an arbitrary spinor. Since it carries $+1$ helicity, we can choose $\mathbf{x}^{h_3}$ to be the wavefunction of particle 3. In the language of Feynman rules, this factor can be viewed as a product of polarization vector and momentum, i.e. $\epsilon\cdot \mathbf{p}_1/\mathbf{m}$. Suppose that $s_1\le s_2$, we begin with the primary rep
\begin{equation}
\mathbf{x}^{h_3}[\mathbf{12}]^{2s_1} [\mathbf{2}|\mathbf{p}_1|\mathbf{2}\rangle^{s_2-s_1},
\end{equation}
and use $T^-_1\cdot T^-_2$ and mass insertion to find other primary reps.

\begin{figure}[htbp]
\centering
\includegraphics[width=0.5\linewidth,valign=c]{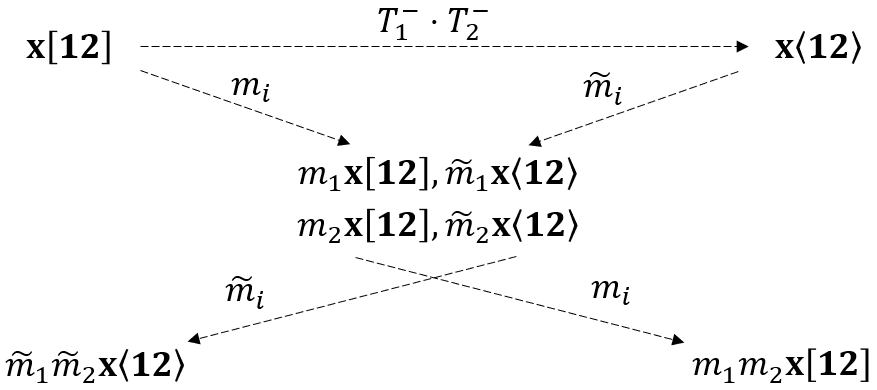}
\caption{The primary and descendant irreps with mass insertion for the $F\bar{F}\gamma$ amplitudes.}
\label{fig:flipFFV}
\end{figure}

In the case $(s_1,s_2,h_3)=(\frac{1}{2},\frac{1}{2},+1)$, all reps are given in figure~\ref{fig:flipFFV}. The general form of $F\bar{F}\gamma$ amplitudes are taken as
\begin{equation} \begin{aligned} \label{eq:FFgamma}
\mathcal{M}(F,\bar{F},\gamma)
=(c_1+c_2 \tilde{m}_1+c_3 \tilde{m}_2+c_4 \tilde{m}_1 \tilde{m}_2)\mathbf{x}\langle\mathbf{12}\rangle
+(c_5+c_6 m_1+c_7 m_2+c_8 m_1 m_2)\mathbf{x}[\mathbf{12}].
\end{aligned} 
\end{equation}
On the other hand, in the AHH formalism since the chirality info is lost, the AHH $F\bar{F}\gamma$ amplitude is thus
\bea
M_{AHH}(\mathbf{1}^{\frac{1}{2}}, \mathbf{2}^{\frac{1}{2}}, 3^{+1})&=g_1\mathbf{x}\langle\mathbf{12}\rangle + g_2 \mathbf{x}^2 \langle\mathbf{1}3\rangle\langle3\mathbf{2}\rangle,
\eea
where the first term corresponds to the minimal coupling, while  the second term are the $g-2$ contributions. The second term can be reduced to
\begin{equation}
\mathbf{x}^2 \langle\mathbf{1}3\rangle\langle3\mathbf{2}\rangle= -\mathbf{x} \langle\mathbf{1}|\mathbf{p}_1|3]\langle3\mathbf{2}\rangle
= \mathbf{x} \langle\mathbf{1}|\mathbf{p}_1(\mathbf{p}_1+\mathbf{p}_2)|\mathbf{2}\rangle
=  \mathbf{m}^2 \mathbf{x}\langle\mathbf{12}\rangle-m_1 m_2\mathbf{x}[\mathbf{12}].
\end{equation}
Compared with Eq.~\eqref{eq:FFgamma}, we find the relation of the coefficients $g_i$ and $c_i$ as
\bea
g_1 &=& (c_1+c_2 \tilde{m}_1+c_3 \tilde{m}_2+c_4 \tilde{m}_1 \tilde{m}_2)+\frac{\mathbf{m}^2}{m_1 m_2}(c_5+c_6 m_1+c_7 m_2+c_8 m_1 m_2),\\
g_2 &=& -\frac{1}{m_1 m_2}(c_5+c_6 m_1+c_7 m_2+c_8 m_1 m_2).
\eea
Comparing to AHH formalism, the advantage of our result in Eq.~\eqref{eq:FFgamma} is that the transversality is explicit, so it has clear UV connection. The AHH amplitude, however, do not care about the transversality, so the UV connection is ambiguous. 

Let us check the UV correspondence of the amplitudes in Eq.~\eqref{eq:FFgamma}. First consider the term $\mathbf{x}\langle\mathbf{12}\rangle$, which gives total transversality $t_1+t_2+t_3=0$. It can corresponds to 4-pt UV massless amplitude in two ways. One way is that we pick the contributions with internal fermion, in which particle 3 couples to fermion. Taking the on-shell limit, which has been used in On-shell Higgsing, we have
\begin{equation}
\lim_{p_4\rightarrow \eta_1}\frac{\langle 1|P_{14}|3]\langle\xi 2\rangle}{\langle\xi 3\rangle s_{14}}-\lim_{p_4\rightarrow \eta_2}\frac{\langle 1\xi\rangle[3|P_{24}|2\rangle}{\langle\xi 3\rangle s_{24}}
=\frac{\langle 1|\eta_1|3]\langle\xi 2\rangle}{\langle\xi 3\rangle \mathbf{m}^2}-\frac{\langle 1\xi\rangle[3|\eta_2|2\rangle}{\langle\xi 3\rangle \mathbf{m}^2}.
\end{equation}
Here we write the $F\bar{F}\gamma S$ amplitude in the form with reference spinor $|\xi\rangle$, so it can match to the $\mathbf{x}$-factor directly. Bolding the massless spinors gives
\begin{equation}
\frac{\langle \mathbf{1}|\mathbf{p}_1|3]\langle\xi \mathbf{2}\rangle}{\langle\xi 3\rangle \mathbf{m}^2}-\frac{\langle \mathbf{1}\xi\rangle[3|\mathbf{p}_2|\mathbf{2}\rangle}{\langle\xi 3\rangle \mathbf{m}^2}
=\frac{\langle \xi|\mathbf{p}_1|3]}{\langle\xi 3\rangle \mathbf{m}^2}\langle\mathbf{12}\rangle
= \frac{1}{\mathbf{m}}\mathbf{x}\langle\mathbf{12}\rangle.
\end{equation}
Another way to give $\mathbf{x}\langle\mathbf{12}\rangle$ is that we pick the contribution with internal scalar boson. In this case, particle 3 couples to the scalar boson, and we should use a different on-shell limit to give the bolding result:
\begin{equation} \label{eq:deform2}
\lim_{p_4\rightarrow p_1}\frac{\langle12\rangle[3|p_4|\xi\rangle}{\langle3\xi\rangle s_{34}}
=\frac{\langle12\rangle[3|p_1|\xi\rangle}{\langle3\xi\rangle s_{31}}
\rightarrow\frac{1}{\mathbf{m}}\langle\mathbf{12}\rangle\mathbf{x}.
\end{equation}
Similarly, the term $\mathbf{x} [\mathbf{12}]$ also corresponds to 4-pt UV amplitude.

Then we can consider $\tilde{m}_2\mathbf{x}\langle\mathbf{12}\rangle$. There are still two ways to give this IR term, depending on which particle couples to particle 3. Figure~\ref{fig:TopDown3} shows that 3-pt UV amplitude as well as 5-pt UV amplitude can match to $\tilde{m}_2\mathbf{x}\langle\mathbf{12}\rangle$, but this matching cannot be realized by 3-pt UV amplitude alone. 
\begin{figure}[htbp]
\centering
\includegraphics[width=0.8\linewidth,valign=c]{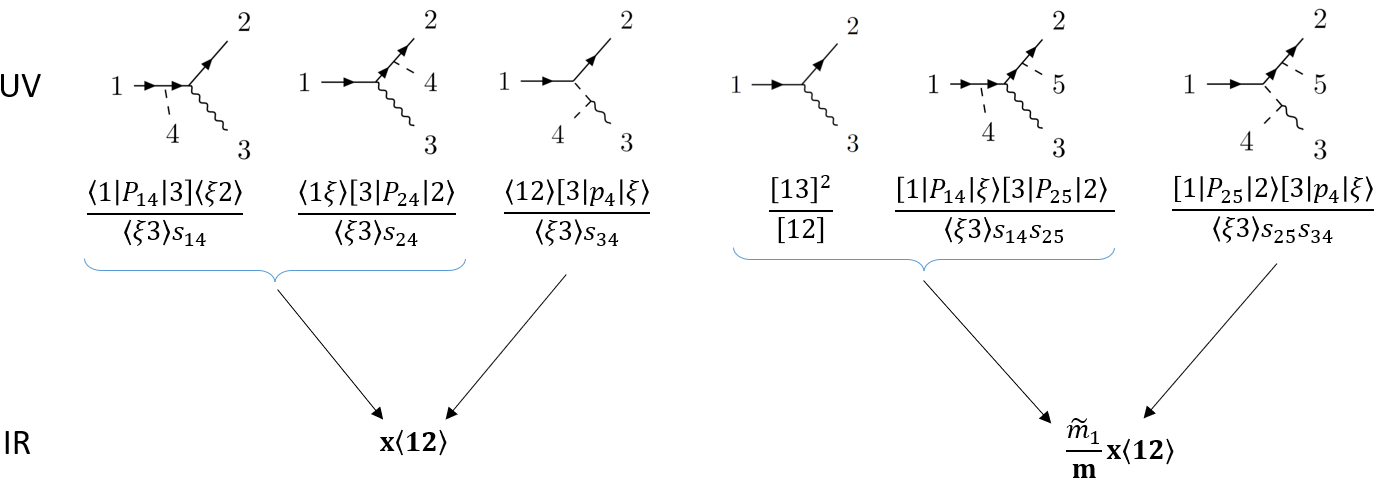}
\caption{The UV massless and IR massive amplitude correspondence for the $c_1$ and $c_2$ structures in the $F\bar{F}\gamma$ amplitudes Eq.~\eqref{eq:FFgamma}. The AHH formalism can only give the Lorentz structure $\mathbf{x}\langle\mathbf{12}\rangle$, while our result tells both the Lorentz structure and its chirality-flip structures. }
\label{fig:TopDown3}
\end{figure}


\section{3. Minimal $F\bar{F}\gamma$ and $e^-e^+ \to \mu^- \mu^+$ in QED}



From Figure~\ref{fig:TopDown3} the IR massive amplitudes still contain unwanted UVs. In QED, we only need certain UV structure. This motivates us match the UV amplitudes to the IR ones at the high energy limit, which help further build the one-to-one correspondence between the UV massless and IR massive amplitude at the high energy limit.


To eliminate the unwanted UV amplitudes, let us consider the high-energy behavior of the 1-massless 2-massive amplitude by using 
\begin{equation}
\begin{cases}
\lambda_{\alpha}^{I} = -\lambda_{\alpha} \zeta^{-I} +\eta_{\alpha} \zeta^{+I}, \\
\tilde{\lambda}_{\dot{\alpha}}^I = \tilde{\lambda}_{\dot{\alpha}} \zeta^{+I} +\tilde{\eta}_{\dot{\alpha}} \zeta^{-I},
\end{cases}
\label{eq:massless-decompose-su2-ahh}
\end{equation}
where $\zeta^{+}$ and $\zeta^{-}$ are two 2-dimensional vectors, $\lambda$ and $\eta$ are the massless spinors with transversality with $\lambda \sim \sqrt{2E}$ while $\eta \sim \frac{\mathbf{m}}{\sqrt{2E}}$ at the high energy limit $E\gg \mathbf{m}$.
In this expansion, the 3-particle kinematics can be converted to 
\begin{equation} \label{eq:kinematic2}
\begin{cases}
    \lambda_1\propto\lambda_2\propto\lambda_3\\
    \tilde{\eta}_1\propto\tilde{\eta}_2\propto\tilde{\lambda}_3
\end{cases}\textrm{or}\quad
\begin{cases}
    \tilde{\lambda}_1\propto\tilde{\lambda}_2\propto\tilde{\lambda}_3\\
    \eta_1\propto\eta_2\propto\lambda_3
\end{cases},
\end{equation}
which is consistent with the equal mass condition $\langle3|\mathbf{p}_1|3]=0$ and momentum conservation. Meanwhile, the $\mathbf{x}$-factor can be decomposed into two new factors: $x$ and $\bar{x}$. In QED, we can choose them to be have transversality $(t_1,t_2,h_3)=(+1,0,+1)$ and $(0,+1,+1)$. So the decomposition is
\begin{equation} \label{eq:decomposeX}
\mathbf{x}=\frac{m_1}{\mathbf{m}}x+\frac{m_2}{\mathbf{m}}\bar{x}\equiv\frac{m_1}{\mathbf{m}}\frac{[3|p_1|\xi\rangle}{m_1 \langle3\xi\rangle}+\frac{m_2}{\mathbf{m}}\frac{[3|\eta_1|\xi\rangle}{m_2 \langle3\xi\rangle},
\end{equation}
The two new factors can not both survive at HE. They are separated by the two kinematics as
\begin{equation} \label{eq:minimal}
\begin{tabular}{c|c}
\makecell{$\lambda_1\propto\lambda_2\propto\lambda_3$\\$\tilde{\eta}_1\propto\tilde{\eta}_2\propto\tilde{\lambda}_3$} & \makecell{$\tilde{\lambda}_1\propto\tilde{\lambda}_2\propto\tilde{\lambda}_3$\\$\eta_1\propto \eta_2\propto\lambda_3$} \\
\hline
$x\neq 0$, $\bar{x}=0$ & $x=0$, $\bar{x}\neq 0$\\
\end{tabular}
\end{equation} 

Using Eq.~\eqref{eq:decomposeX}, the $F\bar{F}\gamma$ amplitudes in Eq.~\eqref{eq:FFgamma} can be further separated into two pieces for two kinematics. At the high energy limit, using the chirality-helicity matching, there is only one helicity structure that satisfy $h_i=t_i$ for each terms in Eq.~\eqref{eq:FFgamma}. 
The selected helicity should be  
\begin{equation} \label{eq:FFgamma2}
\renewcommand{\arraystretch}{1.5}
\begin{tabular}{c|c|c}
helicity & \makecell{$\lambda_1\propto\lambda_2\propto\lambda_3$\\$\tilde{\eta}_1\propto\tilde{\eta}_2\propto\tilde{\lambda}_3$} & \makecell{$\tilde{\lambda}_1\propto\tilde{\lambda}_2\propto\tilde{\lambda}_3$\\$\eta_1\propto \eta_2\propto\lambda_3$} \\
\hline
$(-\frac{1}{2},-\frac{1}{2},+1)$ &
$\frac{c_1}{\mathbf{m}}m_1 x\langle 12\rangle+\frac{c_8}{\mathbf{m}} m_1^2 m_2 x[\eta_1 \eta_2] $  & 
$\frac{c_1}{\mathbf{m}}m_2 \bar{x}\langle 12\rangle+\frac{c_8}{\mathbf{m}} m_1 m_2^2 \bar{x}[\eta_1 \eta_2] $ \\
$(+\frac{1}{2},-\frac{1}{2},+1)$ &
$-c_2 \mathbf{m} x\langle \eta_1 2\rangle+\frac{c_7}{\mathbf{m}} m_1 m_2 x[1 \eta_2] $  & 
$-\frac{c_2}{\mathbf{m}}\tilde{m}_1 m_2 \bar{x}\langle \eta_1 2\rangle+\frac{c_7}{\mathbf{m}} m_2^2 \bar{x}[1 \eta_2] $ \\
$(-\frac{1}{2},+\frac{1}{2},+1)$ &
$-\frac{c_3}{\mathbf{m}}m_1\tilde{m}_2 x\langle 1\eta_2\rangle+\frac{c_6}{\mathbf{m}} m_1^2 x[\eta_1 2] $  & 
$-c_3\mathbf{m} \bar{x}\langle 1\eta_2\rangle+\frac{c_6}{\mathbf{m}} m_1 m_2 \bar{x}[\eta_1 2] $ \\
$(+\frac{1}{2},+\frac{1}{2},+1)$ &
$c_4\mathbf{m}\tilde{m}_2 \bar{x}\langle \eta_1\eta_2\rangle+\frac{c_5}{\mathbf{m}} m_1 \bar{x}[12] $  & 
$c_4 \mathbf{m}\tilde{m}_1 \bar{x}\langle \eta_1 \eta_2\rangle+\frac{c_5}{\mathbf{m}} m_2 \bar{x}[12] $ \\
\end{tabular}
\end{equation}

\begin{figure}[htbp]
\centering
\includegraphics[scale=0.7]{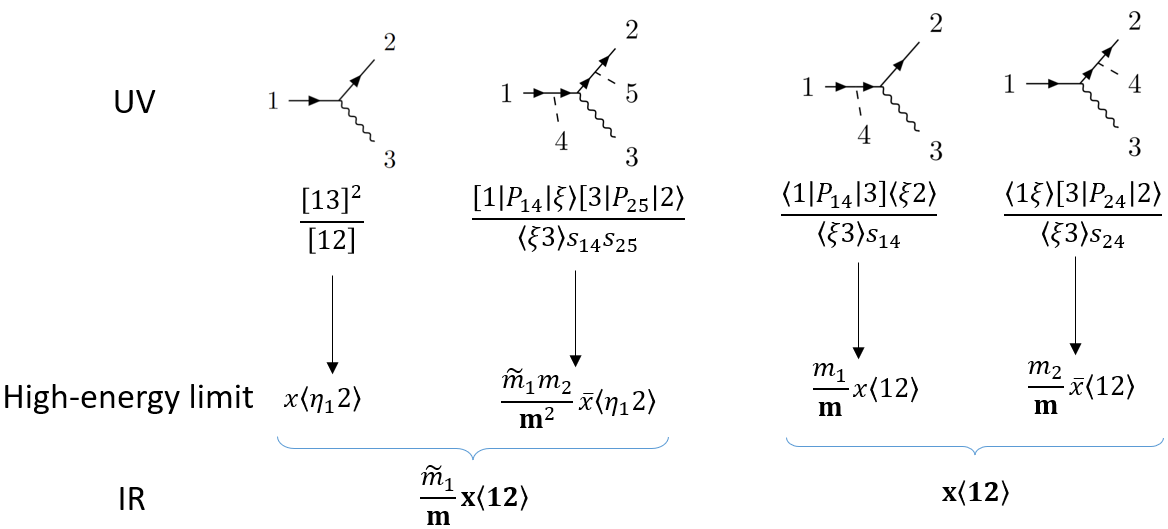}
\caption{The UV massless and IR massive amplitude correspondence for the helicity $(+\frac{1}{2},-\frac{1}{2},+1)$ structure (left two UVs) and $(-\frac{1}{2},-\frac{1}{2},+1)$ structure (right two UVs).}
\label{fig:TopDown4}
\end{figure}

The above Eq.~\eqref{eq:FFgamma2} shows one-to-one correspondence between massless UV and massive IR at the high energy limit, and thus tells us the way that how to isolate different UV and corresponding IR amplitudes using the chirality-helicity matching. Note that each term in Eq.~\eqref{eq:FFgamma2} corresponds to an unique UV amplitude. Let us select UVs in the QED: 
\bit 
\item First the $(-\frac{1}{2},-\frac{1}{2},+1)$ and $(+\frac{1}{2},+\frac{1}{2},+1)$ helicity should be eliminated directly, which tells $c_1=c_8=c_4 = c_5 =0$.
\item Then for $(+\frac{1}{2},-\frac{1}{2},+1)$ and $(-\frac{1}{2},-\frac{1}{2},+1)$ helicity, the UVs of the QED theory should not contain higher dimensional contact amplitudes, which eliminate $c_7 = c_6 =0$. 
\item The only left UVs should be $c_2$ and $c_3$ in the  $(+\frac{1}{2},-\frac{1}{2},+1)$ and $(-\frac{1}{2},-\frac{1}{2},+1)$ helicity.
 The corresponding UV amplitudes are 3-pt and (or) 5-pt amplitudes~\footnote{For some helicity, it is possible that the UV amplitudes contains only 5-pt amplitude. If the 3-pt and 5-pt UVs are both involved, we only take the 3-pt UV in our treatment since 5-pt UV is sub-leading. }, with the on-shell Higgsing
\begin{equation} \begin{aligned}
\lim_{p_4\rightarrow \eta_1,p_5\rightarrow \eta_2}\frac{\langle1|P_{14}|3]\langle\xi|P_{25}|2]}{\langle\xi 3\rangle s_{14}s_{25}}=\frac{\langle1|\eta_1|3]\langle\xi\eta_2\rangle}{\langle\xi 3\rangle m_2\mathbf{m}^2}\rightarrow\frac{1}{\mathbf{m}^2}\bar{x}\langle1\eta_2\rangle,\quad
\frac{[13]^2}{[12]}=\frac{[13]\langle\xi2\rangle}{\langle\xi 3\rangle}\rightarrow x\langle\eta_1 2\rangle.
\end{aligned} \end{equation}

\item Note that the UV amplitude with scalar QED, which is similar to Eq.~\eqref{eq:deform2}, may also contributes the above IR structure. Such UV origins can be dropped by taking only the IR corresponding to either 3-pt or 5-pt UV amplitudes.

\eit


Following this procedure, we select the IR amplitudes $c_2$ and $c_3$ in the high energy limit. The selected amplitudes $F\bar{F}\gamma$ can be written  diagrammatically
\begin{equation} \label{eq:equalFFV1}
\begin{tabular}{c|c|c|c}
helicity & \makecell{$\lambda_1\propto\lambda_2\propto\lambda_3$\\$\tilde{\eta}_1\propto\tilde{\eta}_2\propto\tilde{\lambda}_3$} & helicity & \makecell{$\tilde{\lambda}_1\propto\tilde{\lambda}_2\propto\tilde{\lambda}_3$\\$\eta_1\propto \eta_2\propto\lambda_3$} \\
\hline
$(+\frac{1}{2},-\frac{1}{2},+1)$ & 
\begin{tikzpicture}[baseline=0.7cm] \begin{feynhand}
\vertex [particle] (i1) at (0.15,0.8) {$1^{+}$};
\vertex [particle] (i2) at (0.9+0.7*0.75,0.8+0.8*0.75) {$2^{-}$};
\vertex [particle] (i3) at (0.9+0.7*0.75,0.8-0.8*0.75) {$3^{+}$};
\vertex (v1) at (0.9,0.8);
\vertex [ringdot,cyan] (v3) at (0.9+0.7*0.2,0.8-0.8*0.2) {};
\graph{(i1)--[plain,cyan,very thick] (v1)};
\graph{(i2)--[plain,cyan,very thick] (v1)};
\graph{(i3)--[plain,cyan] (v3)--[plain,cyan] (v1)};
\end{feynhand} \end{tikzpicture}$=-x \langle\eta_1 2\rangle$ & $(-\frac{1}{2},+\frac{1}{2},+1)$ & 
\begin{tikzpicture}[baseline=0.7cm] \begin{feynhand}
\setlength{\feynhanddotsize}{0.8mm}
\vertex [particle] (i1) at (0.15,0.8) {$1^{-}$};
\vertex [particle] (i2) at (0.9+0.7*0.75,0.8+0.8*0.75) {$2^{+}$};
\vertex [particle] (i3) at (0.9+0.7*0.75,0.8-0.8*0.75) {$3^{+}$};
\vertex (v1) at (0.9,0.8);
\vertex [crossdot,cyan] (v3) at (0.9+0.7*0.2,0.8-0.8*0.2) {};
\graph{(i1)--[plain,red,very thick] (v1)};
\graph{(i2)--[plain,red,very thick] (v1)};
\graph{(i3)--[plain,cyan] (v3)--[plain,cyan] (v1)};
\end{feynhand} \end{tikzpicture}$=-\bar{x}\langle 1\eta_2\rangle$ \\
$(-\frac{1}{2},+\frac{1}{2},-1)$ & 
\begin{tikzpicture}[baseline=0.7cm] \begin{feynhand}
\setlength{\feynhanddotsize}{0.8mm}
\vertex [particle] (i1) at (0.15,0.8) {$1^{-}$};
\vertex [particle] (i2) at (0.9+0.7*0.75,0.8+0.8*0.75) {$2^{+}$};
\vertex [particle] (i3) at (0.9+0.7*0.75,0.8-0.8*0.75) {$3^{-}$};
\vertex (v1) at (0.9,0.8);
\vertex [crossdot,red] (v3) at (0.9+0.7*0.2,0.8-0.8*0.2) {};
\graph{(i1)--[plain,red,very thick] (v1)};
\graph{(i2)--[plain,red,very thick] (v1)};
\graph{(i3)--[plain,red] (v3)--[plain,red] (v1)};
\end{feynhand} \end{tikzpicture}$=x^{-1} [\eta_1 2]$ &
$(+\frac{1}{2},-\frac{1}{2},-1)$ & 
\begin{tikzpicture}[baseline=0.7cm] \begin{feynhand}
\vertex [particle] (i1) at (0.15,0.8) {$1^{+}$};
\vertex [particle] (i2) at (0.9+0.7*0.75,0.8+0.8*0.75) {$2^{-}$};
\vertex [particle] (i3) at (0.9+0.7*0.75,0.8-0.8*0.75) {$3^{-}$};
\vertex (v1) at (0.9,0.8);
\vertex [ringdot,red] (v3) at (0.9+0.7*0.2,0.8-0.8*0.2) {};
\graph{(i1)--[plain,cyan,very thick] (v1)};
\graph{(i2)--[plain,cyan,very thick] (v1)};
\graph{(i3)--[plain,red] (v3)--[plain,red] (v1)};
\end{feynhand} \end{tikzpicture}$=\bar{x}^{-1}[1\eta_2]$ \\
\end{tabular}
\end{equation}


For the process $e^-e^+ \to \mu^- \mu^+$ with helicities $(+\frac{1}{2},-\frac{1}{2},+\frac{1}{2},-\frac{1}{2})$, we can glue the 3-pt amplitudes with constraint $h_i=t_i$ to give the correct result. 
In the factorized limit, there are $2\times 2$ possible kinematics, but two of them are enough to give the 4 fermions amplitude. We first choose 
\begin{equation} \begin{aligned} \label{eq:kinematicQED1}
|1\rangle\propto|2\rangle\propto&|P_{12}\rangle\propto|\eta_3\rangle\propto|\eta_4\rangle,\\
|3]\propto|4]\propto&|P_{12}]\propto|\eta_1]\propto|\eta_2].
\end{aligned} \end{equation} 
This kinematics gives two transversality $(\pm\frac{1}{2},\mp\frac{1}{2},\pm\frac{1}{2},\mp\frac{1}{2})$. Diagrammatically, the gluing amplitudes without chirality flip are
\begin{equation} \begin{aligned}
\begin{tikzpicture}[baseline=0.8cm] \begin{feynhand}
\setlength{\feynhandblobsize}{2mm}
\vertex [particle] (i1) at (0.25,1.525) {$1^{+}$};
\vertex [particle] (i2) at (0.2,0.3) {$2^{-}$};
\vertex [particle] (i3) at (2.00,0.3) {$3^{+}$};
\vertex [particle] (i4) at (1.95,1.525) {$4^{-}$};
\vertex (a1) at (0.6-0.6*0.33,0.9+0.9*0.33);
\vertex (a2) at (1.6+0.6*0.33,0.9-0.9*0.33);
\vertex (v1) at (0.6,0.9);
\vertex [ringblob,color=cyan,fill=white] (v2) at (0.9,0.9) {};
\vertex (v5) at (1.1,0.9);
\vertex [ringblob,color=red,fill=white] (v3) at (1.3,0.9) {};
\vertex (v4) at (1.6,0.9);
\graph{(i1)--[plain,cyan,very thick](v1)--[plain,cyan,very thick] (i2)};
\graph{(i4)--[plain,cyan,very thick](v4)--[plain,cyan,very thick](i3)};
\graph{(v1)--[plain,cyan](v2)--[plain,cyan](v5)--[plain,red,slash={[style=black]0}](v3)--[plain,red](v4)};
\end{feynhand} \end{tikzpicture}+
\begin{tikzpicture}[baseline=0.8cm] \begin{feynhand}
\setlength{\feynhanddotsize}{1mm}
\vertex [particle] (i1) at (0.25,1.525) {$1^{+}$};
\vertex [particle] (i2) at (0.2,0.3) {$2^{-}$};
\vertex [particle] (i3) at (2.00,0.3) {$3^{+}$};
\vertex [particle] (i4) at (1.95,1.525) {$4^{-}$};
\vertex (a1) at (0.6-0.6*0.33,0.9+0.9*0.33);
\vertex (a2) at (1.6+0.6*0.33,0.9-0.9*0.33);
\vertex (v1) at (0.6,0.9);
\vertex [crossdot,color=red,fill=white] (v2) at (0.9,0.9) {};
\vertex (v5) at (1.1,0.9);
\vertex [crossdot,color=cyan,fill=white] (v3) at (1.3,0.9) {};
\vertex (v4) at (1.6,0.9);
\graph{(i1)--[plain,red,very thick](v1)--[plain,red,very thick] (i2)};
\graph{(i4)--[plain,red,very thick](v4)--[plain,red,very thick](i3)};
\graph{(v1)--[plain,red](v2)--[plain,red](v5)--[plain,cyan,slash={[style=black]0}](v3)--[plain,cyan](v4)};
\end{feynhand} \end{tikzpicture}.
\end{aligned} \end{equation} 
The numerator of the 4-pt amplitudes, product of two 3-pt amplitudes, is
\begin{equation} \begin{aligned} \label{eq:num1}
&-x_{12}\langle\eta_1 2\rangle\times \bar{x}_{34}^{-1}[3 \eta_4]-x_{12}^{-1}[1\eta_2]\times \bar{x}_{34} \langle\eta_3 4\rangle\\
=&-\frac{[1P_{12}]^2}{[12]}\times\frac{\langle P_{12}4\rangle^2}{\langle 34\rangle}-m_1 \tilde{m}_2\frac{[12]}{[2P_{12}]^2}\times m_3\tilde{m}_4\frac{\langle 34\rangle}{\langle P_{12} 3\rangle^2}\\
=&[13]\langle24\rangle+\frac{m_1 \tilde{m}_2 m_3\tilde{m}_4}{[24]\langle13\rangle}=[13]\langle24\rangle+[\eta_2\eta_4]\langle\eta_1\eta_3\rangle.
\end{aligned} \end{equation}  

Then we consider another kinematics
\begin{equation} \begin{aligned} \label{eq:kinematicQED2}
|1\rangle\propto|2\rangle\propto&|P_{12}\rangle\propto|3\rangle\propto|4\rangle,\\
|\eta_1]\propto|\eta_2]\propto&|P_{12}]\propto|\eta_3]\propto|\eta_4].
\end{aligned} \end{equation} 
It gives the other two transversality $(\pm\frac{1}{2},\mp\frac{1}{2},\mp\frac{1}{2},\pm\frac{1}{2})$ as
\begin{equation} \begin{aligned}
\begin{tikzpicture}[baseline=0.8cm] \begin{feynhand}
\setlength{\feynhandblobsize}{2mm}
\setlength{\feynhanddotsize}{1mm}
\vertex [particle] (i1) at (0.25,1.525) {$1^{+}$};
\vertex [particle] (i2) at (0.2,0.3) {$2^{-}$};
\vertex [particle] (i3) at (2.00,0.3) {$3^{+}$};
\vertex [particle] (i4) at (1.95,1.525) {$4^{-}$};
\vertex (a1) at (0.6-0.6*0.33,0.9+0.9*0.33);
\vertex (a2) at (1.6+0.6*0.33,0.9+0.9*0.33);
\vertex (v1) at (0.6,0.9);
\vertex [ringblob,color=cyan,fill=white] (v2) at (0.9,0.9) {};
\vertex (v5) at (1.1,0.9);
\vertex [crossdot,color=red,fill=white] (v3) at (1.3,0.9) {};
\vertex (v4) at (1.6,0.9);
\graph{(i1)--[plain,cyan,very thick](v1)--[plain,cyan,very thick] (i2)};
\graph{(i4)--[plain,red,very thick](v4)--[plain,red,very thick](i3)};
\graph{(v1)--[plain,cyan](v2)--[plain,cyan](v5)--[plain,red,slash={[style=black]0}](v3)--[plain,red](v4)};
\end{feynhand} \end{tikzpicture}+
\begin{tikzpicture}[baseline=0.8cm] \begin{feynhand}
\setlength{\feynhandblobsize}{2mm}
\setlength{\feynhanddotsize}{1mm}
\vertex [particle] (i1) at (0.25,1.525) {$1^{+}$};
\vertex [particle] (i2) at (0.2,0.3) {$2^{-}$};
\vertex [particle] (i3) at (2.00,0.3) {$3^{+}$};
\vertex [particle] (i4) at (1.95,1.525) {$4^{-}$};
\vertex (a1) at (0.6-0.6*0.33,0.9+0.9*0.33);
\vertex (a2) at (1.6+0.6*0.33,0.9+0.9*0.33);
\vertex (v1) at (0.6,0.9);
\vertex [crossdot,color=red,fill=white] (v2) at (0.9,0.9) {};
\vertex (v5) at (1.1,0.9);
\vertex [ringblob,color=cyan,fill=white] (v3) at (1.3,0.9) {};
\vertex (v4) at (1.6,0.9);
\graph{(i1)--[plain,red,very thick](v1)--[plain,red,very thick] (i2)};
\graph{(i4)--[plain,cyan,very thick](v4)--[plain,cyan,very thick](i3)};
\graph{(v1)--[plain,red](v2)--[plain,red](v5)--[plain,cyan,slash={[style=black]0}](v3)--[plain,cyan](v4)};
\end{feynhand} \end{tikzpicture}.
\end{aligned} \end{equation} 
Summing over transversality, we obtain the numerator of the 4-pt amplitudes
\begin{equation} \begin{aligned} \label{eq:num2}
&-x_{12}\langle\eta_1 2\rangle\times x_{34}^{-1}[3 \eta_4]-x_{12}^{-1}[1\eta_2]\times x_{34} \langle\eta_3 4\rangle\\
=&\frac{[1P_{12}]^2}{[12]}\times m_3 \tilde{m}_4\frac{[34]}{[P_{12}4]^2}+m_1 \tilde{m}_2\frac{[12]}{[2P_{12}]^2}\times \frac{[P_{12}3]^2}{[34]}\\
=&-[1\eta_4]\langle2\eta_3\rangle-[\eta_2 3]\langle\eta_1 4\rangle.
\end{aligned} \end{equation} 

Bolding the spinors in Eqs.~\eqref{eq:num1} and \eqref{eq:num2} and adding the pole structure, we get the final result
\begin{equation} \begin{aligned}
\mathcal{M}({e\bar{e}\to\mu\bar{\mu}})=\frac{[\mathbf{13}]\langle\mathbf{24}\rangle+[\mathbf{24}]\langle\mathbf{13}\rangle+[\mathbf{14}]\langle\mathbf{23}\rangle+[\mathbf{23}]\langle\mathbf{14}\rangle}{\mathbf{s}_{12}}.
\end{aligned} \end{equation}

\section{4. Young Diagram Representation of the $SO(5,1)$ Group}

We present the Young diagram method to systematically describe the $(\Delta,J_1,J_2)$ irrep of the $SO(5,1)$ group, as shown in Eq.~\eqref{eq:spin-halfso6} and ~\eqref{eq:spin-1so6}.
The generic irrep of the $SO(5,1)$ is labeled as $(\Delta, J_1, J_2)$, while the LG of the $ISO(5,1)$ is $SU(2) \times SU(\overline{2})$. Generally, the $(\Delta,J_1,J_2)$ irrep is divided into $SU(2) \times SU(\overline{2})$ LG invariant and covariant structures. The LG invariant structure can be obtained by the $SU(4)$ Young diagram 
\newline\newline
\eq{
\arraycolsep=0pt\def\arraystretch{0}
\begin{array}{ccc}
\overmat{\Delta-J_1-J_2}{\ytableausetup{centertableaux}
\ydiagram[*(white) \bullet]{1,1} & \cdots & \ydiagram[*(white) \bullet]{1,1}}
\end{array}
\sim \frac{1}{(\Delta-J_1-J_2)!} \left( p_{A_1 B_1} \cdots p_{A_{\Delta-J_1-J_2} B_{\Delta-J_1-J_2}} +perm.(A_i) \right),
}
where $\ydiagram[*(white) \bullet]{1,1} \sim p_{AB}$ denotes a 6D momentum.
On the other hand, the $SU(2)$ and $SU(\overline{2})$ covariant structures are
\bea
\arraycolsep=0pt\def\arraystretch{0}
\begin{array}{ccc}
\undermat{2J_2}{\ydiagram[*(orange)]{1} & \cdots & \ydiagram[*(orange)]{1}}
\end{array}
\sim \lambda_{A_1}^{(I_1} \cdots \lambda_{A_{J_2}}^{I_{2J_2})} , \\
\nonumber \\
\nonumber \\
\arraycolsep=0pt\def\arraystretch{0}
\begin{array}{ccc}
\undermat{2J_1}{\ydiagram[*(blue)]{1,1,1} & \cdots & \ydiagram[*(blue)]{1,1,1}}
\end{array}
\sim \tilde{\lambda}_{A_1 B_1 C_1}^{(\overline{I}_1} \cdots \tilde{\lambda}_{A_{2J_1} B_{2J_1} C_{2J_1}}^{\overline{I}_{J_1})},
\eea\newline
where $\ydiagram[*(orange)]{1} \sim \lambda_A^{I}$ denotes the 6D spinor of $SU(2)$, and $\ydiagram[*(blue)]{1,1,1} \sim \epsilon_{ABCD}\tilde{\lambda}^{D \overline{I}} \equiv \tilde{\lambda}_{ABC}^{\overline{I}} $ denotes the 6D spinor of $SU(\overline{2})$. 
The direct product of LG invariant and covariant $SU(4)$ Young diagrams gives
\newline\newline
\eq{
& \arraycolsep=0pt\def\arraystretch{0}
\begin{array}{ccc}
\overmat{\Delta-J_1 -J_2}{\ydiagram[*(white) \bullet]{1,1} & \cdots & \ydiagram[*(white) \bullet]{1,1}}
\end{array}
\otimes
\arraycolsep=0pt\def\arraystretch{0}
\begin{array}{ccc}
\undermat{2J_1}{\ydiagram[*(blue)]{1,1,1} & \cdots & \ydiagram[*(blue)]{1,1,1}}
\end{array}
\otimes
\arraycolsep=0pt\def\arraystretch{0}
\begin{array}{ccc}
\undermat{2J_2}{\ydiagram[*(orange)]{1} & \cdots & \ydiagram[*(orange)]{1}}
\end{array} = 
\arraycolsep=0pt\def\arraystretch{0}
\begin{array}{ccccccccc}
\overmat{2J_1}{\ydiagram[*(blue)]{1} & \cdots & \ydiagram[*(blue)]{1}} & \ydiagram[*(white) \bullet]{1} & \cdots & \ydiagram[*(white) \bullet]{1} & \overmat{2J_2}{\ydiagram[*(orange)]{1} & \cdots & \ydiagram[*(orange)]{1}} \\
\ydiagram[*(blue)]{1} & \cdots & \ydiagram[*(blue)]{1} & \undermat{\Delta-J_1-J_2}{\ydiagram[*(white) \bullet]{1} & \cdots & \ydiagram[*(white) \bullet]{1}} &&& \\
\ydiagram[*(blue)]{1} & \cdots & \ydiagram[*(blue)]{1} &&&&&&
\end{array} \\
& \\
\sim & \tilde{\lambda}_{A_1 B_1 C_1}^{\overline{I}_1} \cdots \tilde{\lambda}_{A_{2J_1} B_{2J_1} C_{2J_1}}^{\overline{I}_{2J_1}} p_{A_{2J_1+1} B_{2J_1+1}} \cdots p_{A_{\Delta+J_1-J_2} B_{\Delta+J_1-J_2}} \lambda_{A_{\Delta+J_1-J_2+1}}^{I_{1}} \cdots \lambda_{A_{\Delta+J_1+2J_2}}^{I_{2J_2}} +perm.
}
where the Young diagram of the decedent reps vanishes due to $\tilde{\lambda}^{A\overline{I}} \lambda_{A}^{I}=0$.
The spinor indices $A$, $B$ and $C$ should be symmetrized, which is guaranteed by the Young tableaux of the permutation group $S_{\Delta +J_1 +J_2}$. This primary Young diagram constitutes the general $SO(5,1)$ spinor.

In this work, we focus on the reduced $SU(2)$ LG, which corresponds to the symmetry breaking $ISO(5,1) \rightarrow ISO(2) \times ISO(3,1)$. When a subgroup $SU(2) \subseteq SU(2) \times SU(\overline{2})$ is considered, a subspace of the complete set of irreps, $\bigoplus_{\Delta, J_1, J_2} (\Delta, J_1, J_2)$ is obtained. Consequently, the relevant $SO(5,1)$ irreps are labeled as $(\Delta, 0, J_2)$, with the corresponding $SU(4)$ Young diagram 
\eq{\label{eq:d-0-j-so6}
(\Delta,0,J_2) \sim \ 
\arraycolsep=0pt\def\arraystretch{0}
\begin{array}{cccccc}
\ydiagram[*(white) \bullet]{1} & \cdots & \ydiagram[*(white) \bullet]{1} & \undermat{2s}{\ydiagram[*(orange)]{1} & \cdots & \ydiagram[*(orange)]{1}} \\
\undermat{\Delta-s}{\ydiagram[*(white) \bullet]{1} & \cdots & \ydiagram[*(white) \bullet]{1}} &&&
\end{array} \sim& \lambda_{A_1}^{(I_1} \cdots \lambda_{A_{\Delta+s}}^{I_{\Delta+s})} \lambda_{B_{1} I_1} \cdots \lambda_{B_{\Delta-s} I_{\Delta-s}} \\
=& \frac{1}{(\Delta+s)!}\left(p_{A_1 B_1} \cdots p_{A_{\Delta-s} B_{\Delta-s}} \lambda_{A_{\Delta-s+1}}^{I_{\Delta-s+1}} \cdots \lambda_{A_{\Delta+s}}^{I_{\Delta+s}} +perm.(A_i)\right),
}
where $s=J_2$ labels the spin of the reduced $SU(2)$ LG.

The $(\Delta,0,J_2)$ irrep is obtained by the subduced representation
\eq{
(\Delta,0,J_2) = \bigoplus_{t,j_1,j_2} [t,j_1,j_2],
}
where $[t,j_1,j_2]$ is an $SO(2) \times SO(3,1)$ irrep, and the highest weight is $[t_{\max}, 0, s] = [\Delta, 0, J_2]$. 
Each $[t,j_1,j_2]$ irrep is obtained by repeatedly acting the $T^{-}$ operator on $[t+1,j'_1,j'_2]$ from the highest weight $[t_{\max}, 0, s]$. 

For $SO(2) \times SO(3,1)$ group, the $[t,j_1,j_2]$ irreps can be written as the $SU(2)$ Young diagram 
\eq{\label{eq:t-j-j-so2}
[t,j_1,j_2] =& \left[\frac{\tilde{n}-n}{2}+\tilde{n}'-n', \frac{n+n_p}{2}, \frac{\tilde{n}+n_p}{2} \right] \\ & \\
\sim & \ 
\arraycolsep=0pt\def\arraystretch{0}
\begin{array}{ccccccccc}
{\color{orange}\yng(1)} & \cdots & {\color{orange}\yng(1)} & {\color{orange}\yng(1)} & \cdots & {\color{orange}\yng(1)} & \yng(1) & \cdots & \yng(1) \\
\undermat{\tilde{n}'}{ {\color{orange}\yng(1)} & \cdots & {\color{orange}\yng(1)}} & \undermat{n_p}{ \yng(1) & \cdots & \yng(1)} & \undermat{n'}{\yng(1) & \cdots & \yng(1)}
\end{array} 
\otimes \arraycolsep=0pt\def\arraystretch{0}
\begin{array}{cccccc}
\overmat{\tilde{n}}{{\color{orange} \yng(1)} & \cdots & {\color{orange} \yng(1)}} & \overmat{n}{\yng(1) & \cdots & \yng(1)}
\end{array} \\
&\\
\sim& \tilde{m}^{\tilde{n}'} m^{n'} \mathbf{p}^{n_p} \tilde{\lambda}_{\dot{\alpha}_1}^{(I_1} \cdots \tilde{\lambda}_{\dot{\alpha}_{\tilde{n}}}^{I_{\tilde{n}_1}} \lambda_{\alpha_1}^{I_{\tilde{n}+1}} \cdots \lambda_{\alpha_{n}}^{I_{\tilde{n}+n})} ,
}
where ${\color{orange}\yng(1)} \sim \tilde{\lambda}_{\dot{\alpha}}^{I}, \yng(1) \sim \lambda_{\alpha}^{I} $ denote a 4D spinor respectively, and the orange boxes are always placed to the left of the black boxes. 
The primary rep corresponds to $(\tilde{n}',n_p,n') = (0,0,0)$.
%
As a subspace of $(\Delta,0,J_2)$, the Young diagram of $[t,j_1,j_2]$ is in the same shape as the one of $(\Delta,0,J_2)$ in Eq.~\eqref{eq:d-0-j-so6} after direct product of the two Young diagram blocks in Eq.~\eqref{eq:t-j-j-so2}, so 
\eq{
\tilde{n}+n= 2J_2, \
\tilde{n}'+n_p+n'= \Delta-J_2.
}
Diagrammatically, the $T^{\pm}$ generator acting on a generic Young diagram gives\newline\newline
\eq{
T^{+} \ \arraycolsep=0pt\def\arraystretch{0}
\begin{array}{ccccccccccc}
\overmat{\tilde{n}_1}{ {\color{orange}\yng(1)} & \cdots & {\color{orange}\yng(1)} & {\color{orange}\yng(1)} & \cdots & {\color{orange}\yng(1)} } & \overmat{n_1}{ \yng(1) & \cdots & \yng(1) & \cdots & \yng(1)} \\
\undermat{\tilde{n}_2}{ {\color{orange}\yng(1)} & \cdots & {\color{orange}\yng(1)}} & \undermat{n_2}{ \yng(1) & \cdots & \yng(1) & \yng(1) & \cdots & \yng(1)} &&
\end{array} 
= \begin{cases}
\arraycolsep=0pt\def\arraystretch{0}
\begin{array}{ccccccccccc}
\overmat{\tilde{n}_1+1}{ {\color{orange}\yng(1)} & \cdots & {\color{orange}\yng(1)} & {\color{orange}\yng(1)} & \cdots & {\color{orange}\yng(1)} } & \overmat{n_1-1}{ \yng(1) & \cdots & \yng(1) & \cdots & \yng(1)} \\
\undermat{\tilde{n}_2}{ {\color{orange}\yng(1)} & \cdots & {\color{orange}\yng(1)}} & \undermat{n_2}{ \yng(1) & \cdots & \yng(1) & \yng(1) & \cdots & \yng(1)} &&
\end{array} \\ \\ \\
\arraycolsep=0pt\def\arraystretch{0}
\begin{array}{ccccccccccc}
\overmat{\tilde{n}_1}{ {\color{orange}\yng(1)} & \cdots & {\color{orange}\yng(1)} & {\color{orange}\yng(1)} & \cdots & {\color{orange}\yng(1)} } & \overmat{n_1}{ \yng(1) & \cdots & \yng(1) & \cdots & \yng(1)} \\
\undermat{\tilde{n}_2+1}{ {\color{orange}\yng(1)} & \cdots & {\color{orange}\yng(1)}} & \undermat{n_2-1}{ \yng(1) & \cdots & \yng(1) & \yng(1) & \cdots & \yng(1)} &&
\end{array}
\end{cases}, \\ \\ \\
T^{-} \ \arraycolsep=0pt\def\arraystretch{0}
\begin{array}{ccccccccccc}
\overmat{\tilde{n}_1}{ {\color{orange}\yng(1)} & \cdots & {\color{orange}\yng(1)} & {\color{orange}\yng(1)} & \cdots & {\color{orange}\yng(1)} } & \overmat{n_1}{ \yng(1) & \cdots & \yng(1) & \cdots & \yng(1)} \\
\undermat{\tilde{n}_2}{ {\color{orange}\yng(1)} & \cdots & {\color{orange}\yng(1)}} & \undermat{n_2}{ \yng(1) & \cdots & \yng(1) & \yng(1) & \cdots & \yng(1)} &&
\end{array} 
= \begin{cases}
\arraycolsep=0pt\def\arraystretch{0}
\begin{array}{ccccccccccc}
\overmat{\tilde{n}_1-1}{ {\color{orange}\yng(1)} & \cdots & {\color{orange}\yng(1)} & {\color{orange}\yng(1)} & \cdots & {\color{orange}\yng(1)} } & \overmat{n_1+1}{ \yng(1) & \cdots & \yng(1) & \cdots & \yng(1)} \\
\undermat{\tilde{n}_2}{ {\color{orange}\yng(1)} & \cdots & {\color{orange}\yng(1)}} & \undermat{n_2}{ \yng(1) & \cdots & \yng(1) & \yng(1) & \cdots & \yng(1)} &&
\end{array} \\ \\ \\
\arraycolsep=0pt\def\arraystretch{0}
\begin{array}{ccccccccccc}
\overmat{\tilde{n}_1}{ {\color{orange}\yng(1)} & \cdots & {\color{orange}\yng(1)} & {\color{orange}\yng(1)} & \cdots & {\color{orange}\yng(1)} } & \overmat{n_1}{ \yng(1) & \cdots & \yng(1) & \cdots & \yng(1)} \\
\undermat{\tilde{n}_2-1}{ {\color{orange}\yng(1)} & \cdots & {\color{orange}\yng(1)}} & \undermat{n_2+1}{ \yng(1) & \cdots & \yng(1) & \yng(1) & \cdots & \yng(1)} &&
\end{array}
\end{cases}.
}

\end{widetext}

\end{document}